\documentclass[11pt]{cernrep}
\usepackage{graphicx,epsfig}
\newcommand{\be}{\begin{equation}}
\newcommand{\ee}{\end{equation}}
\newcommand{\bea}{\begin{eqnarray}}
\newcommand{\eea}{\end{eqnarray}}

\newcommand{\as}{$\alpha_s$}
\def\lsim{\mathrel{\mathpalette\@versim<}}
\def\gsim{\mathrel{\mathpalette\@versim>}}
 \def\@versim#1#2{\lower0.2ex\vbox{\baselineskip\z@skip\lineskip\z@skip
       \lineskiplimit\z@\ialign{$\m@th#1\hfil##$\crcr#2\crcr\sim\crcr}}}
\catcode`@=12 

\newcommand{\ben}{\begin{enumerate}}
\newcommand{\een}{\end{enumerate}}
\newcommand{\la}{\left\langle}
\newcommand{\ra}{\right\rangle}
\newcommand{\lc}{\left[}
\newcommand{\rc}{\right]}
\newcommand{\lp}{\left(}
\newcommand{\rp}{\right)}

\def\frac#1#2{{{#1}\over {#2}}}
\def\gsim{\mathrel{\rlap{\lower4pt\hbox{\hskip1pt$\sim$}}
    \raise1pt\hbox{$>$}}}         
\def\lsim{\mathrel{\rlap{\lower4pt\hbox{\hskip1pt$\sim$}}
    \raise1pt\hbox{$<$}}}         

\newcommand{\rep}{\mathrm{rep}}

\newcommand{\draft}[1]{}

\def\nn{\nonumber}

\def \n0{N_j^{(0)}}

\def\lapprox{\lower .7ex\hbox{$\;\stackrel{\textstyle <}{\sim}\;$}}
\def\gapprox{\lower .7ex\hbox{$\;\stackrel{\textstyle >}{\sim}\;$}}

\begin{document}

\title{\boldmath The PDF4LHC Working Group Interim Report}


\author{
Sergey Alekhin$^{1,2}$,
Simone Alioli$^1$,
Richard ~D. ~Ball$^3$,
Valerio~Bertone$^4$,
Johannes Bl\"umlein$^1$,
Michiel Botje$^5$,
Jon Butterworth$^6$,
Francesco~Cerutti$^7$,
Amanda Cooper-Sarkar$^8$,
Albert de Roeck$^9$,
Luigi~Del~Debbio$^3$,
Joel Feltesse$^{10}$,
Stefano Forte$^{11}$,
Alexander Glazov$^{12}$,
Alberto~Guffanti$^4$,
Claire Gwenlan$^8$,
Joey Huston$^{13}$,
Pedro Jimenez-Delgado$^{14}$,
Hung-Liang Lai$^{15}$,
Jos\'e~I.~Latorre$^7$,
Ronan McNulty$^{16}$,
Pavel Nadolsky$^{17}$,
Sven Olaf Moch$^1$,
Jon Pumplin$^{13}$,
Voica Radescu$^{18}$,
Juan~Rojo$^{11}$,
Torbj\"orn Sj\"ostrand$^{19}$,
W.J. Stirling$^{20}$,
Daniel Stump$^{13}$,
Robert ~S.~ Thorne$^6$,
Maria~Ubiali$^{21}$,
Alessandro Vicini$^{11}$,
Graeme Watt$^{22}$, 
C.-P. Yuan$^{13}$
}


\institute{
$^1$ Deutsches Elektronen-Synchrotron, DESY, Platanenallee 6, D-15738 Zeuthen, Germany \\
$^2$ Institute for High Energy Physics, IHEP, Pobeda 1, 142281 Protvino, Russia\\
$^3$ School of Physics and Astronomy, University of Edinburgh, JCMB, KB, Mayfield Rd, Edinburgh EH9 3JZ, Scotland\\
$^4$ Physikalisches Institut, Albert-Ludwigs-Universit\"at Freiburg, Hermann-Herder-Stra\ss e 3, D-79104 Freiburg i. B., Germany  \\
$^5$ NIKHEF, Science Park, Amsterdam, The Netherlands\\
$^6$ Department of Physics and Astronomy, University College, London, WC1E 6BT, UK\\
$^7$ Departament d'Estructura i Constituents de la Mat\`eria, Universitat de Barcelona, Diagonal 647, E-08028 Barcelona, Spain\\
$^8$ Department of Physics, Oxford University, Denys Wilkinson Bldg, Keble Rd, Oxford, OX1 3RH, UK\\
$^9$ CERN, CH--1211 Gen\`eve 23, Switzerland; Antwerp University, B--2610 Wilrijk, Belgium; University of California Davis, CA, USA\\
$^{10}$ CEA, DSM/IRFU, CE-Saclay, Gif-sur-Yvetee, France\\
$^{11}$ Dipartimento di Fisica, Universit\`a di Milano and INFN, Sezione di Milano, Via Celoria 16, I-20133 Milano, Italy\\
$^{12}$ Deutsches Elektronensynchrotron DESY Notkestra{\ss}e 85 D--22607 Hamburg, Germany \\
$^{13}$ Physics and Astronomy Department, Michigan State University, East Lansing, MI 48824, USA\\
$^{14}$ Institut f{\"u}r Theoretische Physik, 
  Universit{\"a}t Z{\"u}rich, CH-8057 Z{\"u}rich, Switzerland\\
$^{15}$ Taipei Municipal University of Education, Taipei, Taiwan\\
$^{16}$ School of Physics, University College Dublin Science Centre North, UCD Belfeld, Dublin 4, Ireland\\
$^{17}$ Department of Physics, Southern Methodist University, Dallas, TX 75275-0175, USA\\
$^{18}$ Physikalisches Institut, Universit\"at Heidelberg Philosophenweg 12, D--69120 Heidelberg, Germany \\
$^{19}$ Department of Astronomy and Theoretical Physics, Lund University, S\"olvegatan 14A, S-223 62 Lund, Sweden\\
$^{20}$ Cavendish Laboratory, University of Cambridge, CB3 OHE, UK\\
$^{21}$ Institut f\"ur Theoretische Teilchenhysik und Kosmologie, RWTH Aachen University, D-52056 Aachen, Germany\\
$^{22}$ Theory Group, Physics Department, CERN, CH-1211 Geneva 23, Switzerland
}

\maketitle

\clearpage

\setcounter{page}{2}

\begin{abstract}

This document is intended as a study of benchmark cross sections at the LHC (at 7 TeV) at NLO using modern PDFs currently available from the 6 PDF fitting groups that have participated in this exercise. It also contains a succinct user guide to the
computation of PDFs, uncertainties and correlations using available
PDF sets.

A companion note provides an interim summary of the current
recommendations of the PDF4LHC working group for the use of parton
distribution functions (PDFs) and of PDF uncertainties at the LHC, for cross
section and cross section uncertainty calculations. 


\end{abstract}

\clearpage

\tableofcontents

\clearpage

\section{Introduction}
\label{intro}

The LHC experiments are currently producing cross sections from the 7 TeV data, and thus need accurate predictions for these cross sections and their uncertainties at NLO and NNLO. Crucial to the predictions and their uncertainties are the parton distribution functions (PDFs) obtained 
from global fits to data from deep-inelastic scattering, Drell-Yan and
jet data. A number of groups have produced publicly available PDFs
using different data sets and analysis frameworks. It is one of the
charges of the PDF4LHC working group to evaluate and understand
differences among the PDF sets to be used at the LHC, and to provide a
protocol for both experimentalists and theorists to use the PDF sets
to calculate central cross sections at the LHC, as well as to estimate
their PDF uncertainty. This current note is intended to be an interim
summary of our level of understanding of NLO predictions as the first LHC cross sections
at 7 TeV are being produced~\footnote{Comparisons at NNLO for $W$,$Z$ and Higgs production can be found in ref.~\cite{Alekhin:2010dd}}. The intention is to modify this note as 
improvements in data/understanding warrant.

For the purpose of increasing our quantitative understanding of
the similarities and differences between available PDF
determinations, a benchmarking exercise between the different groups
was performed. This exercise 
was very instructive in understanding many differences in
 the PDF analyses: different input data, different methodologies and criteria
 for 
determining uncertainties, 
different ways of 
parametrizing PDFs, different
number of parametrized PDFs, different treatments of heavy quarks, 
different perturbative orders,
different ways of treating  $\alpha_s$ (as an input or as a fit
parameter),
different values of physical parameters such as $\alpha_s$ itself and
heavy quark masses, and more. 
This exercise was also very
instructive in understanding where the PDFs agree and where they
disagree: it  established a 
broad agreement of PDFs (and uncertainties) obtained from data sets of
comparable size and it singled out relevant instances of disagreement
and of dependence of the results on assumptions or methodology.

The outline of this interim report is as follows. 
The first three sections are devoted to a description of current PDF
sets and their usage.
In Sect.~\ref{sec:pdfdet1} we present
several modern PDF determinations, with special regard to the way PDF
uncertainties are determined. First we summarize the main features of
various sets, then we provide an explicit users' guide for the
computation of PDF uncertainties. In Sect.~\ref{sec:pdfdet2} we
discuss theoretical uncertainties on PDFs. We first introduce various
theoretical uncertainties, then we focus on the uncertainty related to
the strong coupling and also in this case we give both a presentation
of choices made by different groups and a users' guide for the
computation of combined PDF+$\alpha_s$ uncertainties. Finally in
Sect.~\ref{sec:corr} we discuss PDF correlations and the way they can
be computed. 

In Sect.~\ref{sec:benchmarking} we introduce the settings for the PDF4LHC
benchmarks on LHC observables, present the results from the different groups
and compare their predictions for important LHC observables at 7 TeV at NLO.
In Sect.~\ref{sec:summary}
we conclude and briefly discuss prospects for future developments.

\clearpage

\section{PDF determinations - experimental uncertainties}
\label{sec:pdfdet1}

Experimental uncertainties of PDFs
 determined in global fits (usually called ``PDF uncertainties'' for
 short)  
reflect three aspects of
 the analysis, and differ because of different choices made in each of
 these aspects: (1) the choice of data set; (2) the 
type of uncertainty estimator used which is used to determine the uncertainties and which also
determines the way in which PDFs are delivered to the user; (3) the
form and size of parton parametrization. First, we briefly discuss
the available options for each of these aspects (at least, those which
have been explored by the various groups discussed here) and summarize
the choices made by each group; then, we provide a concise user guide
for the determination of PDF uncertainties for available fits.
We will in particular discuss the following PDF sets (when several
releases are available the most recent
published ones are given in parenthesis in each case):
ABKM/ABM~\cite{Alekhin:2009ni, Alekhin:2010iu}, CTEQ/CT (CTEQ6.6~\cite{Nadolsky:2008zw}, CT10~\cite{Lai:2010vv}),
GJR~\cite{Gluck:2007ck,Gluck:2008gs}, HERAPDF (HERAPDF1.0~\cite{herapdf10}), MSTW (MSTW08~\cite{Martin:2009iq}), NNPDF (NNPDF2.0~\cite{Ball:2010de}). There is a significant time-lag between the development of a new PDF and the wide adoption of its use by experimental collaborations, so in some cases, we report not on the most up-to-date PDF from a particular group, but instead on the most widely-used.

\subsection{Features, tradeoffs and choices}
\label{ftc}

\subsubsection{Data Set}
\label{data}
There is a clear tradeoff between the size and the consistency of a
data set: a wider data set contains more information, but data coming
from different experiment may be inconsistent to some extent. The
choices made by the various groups are the following:
\begin{itemize}
\item The CTEQ, MSTW and NNPDF data sets considered here include
 both electroproduction and hadroproduction data, in each case both from
  fixed-target and collider experiments. The electroproduction data
  include electron, muon and neutrino  
 deep--inelastic scattering data (both inclusive and charm
 production). The hadroproduction data include Drell-Yan (fixed target
 virtual photon and collider $W$ and $Z$ production) and jet
 production~\footnote{Although the comparisons included in this note are only at NLO,we note that, to date, the inclusive jet cross section, unlike the other processes in the list above, has been calculated only to NLO, and not to NNLO. This may have an impact on the precision of NNLO global PDF fits that include inclusive jet data.}.
\item The GJR data set includes electroproduction data from
  fixed-target and collider experiments, and a smaller set of 
hadroproduction data. The electroproduction data
  include electron and muon  inclusive
 deep--inelastic scattering data, and deep-inelastic charm 
 production from charged leptons and neutrinos. The hadroproduction
 data includes fixed--target virtual photon Drell-Yan production and
 Tevatron jet production.
\item The ABKM/ABM  data sets include
  electroproduction  from
  fixed-target and collider experiments, and fixed--target
  hadroproduction data. The electroproduction data
  include electron, muon and neutrino  
 deep--inelastic scattering data (both inclusive and charm
 production). The hadroproduction data include fixed--target virtual
 photon Drell-Yan production. The most recent version, ABM10~\cite{Trento}, includes Tevatron jet data.
\item The HERAPDF data set  includes all HERA deep-inelastic inclusive
data.  
\end{itemize}

\subsubsection{Statistical treatment}
\label{stat}

Available PDF determinations fall in two broad categories: those
based on a Hessian approach and those which use a Monte Carlo
approach. The delivery of PDFs is different in each case and will be
discussed in Sect.~\ref{delusage}.

Within the Hessian method, PDFs are determined by minimizing a
suitable log-likelihood $\chi^2$ function. Different groups may use somewhat different definitions of $\chi^2$, for example, by including entirely, or only partially, correlated systematic uncertainties. While some groups account for correlated uncertainties by means of a covariance matrix, other groups treat some correlated systematics (specifically but not exclusively normalization uncertanties) as a shift of data, with a penalty term proportional to some power of the shift parameter added to the $\chi^2$. The reader is referred to the original papers for the precise definition adopted by each group, but it should be born in mind that because of all these differences,  
values of the $\chi^2$
quoted by different groups are in general only roughly comparable.

\begin{figure}
\begin{center}
\includegraphics[width=0.60\textwidth]{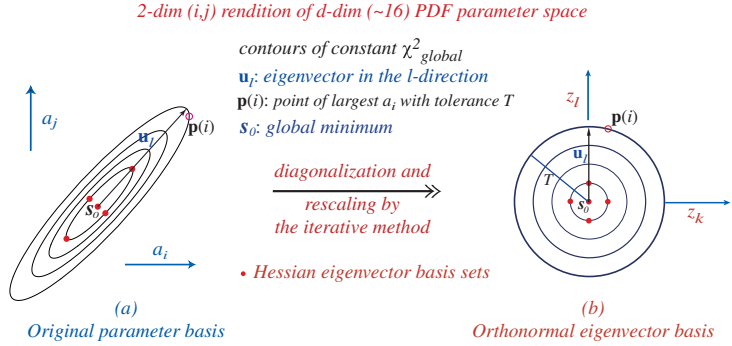}
\caption{\label{fig:hessian} A schematic representation of the transformation from the PDF parameter basis to the orthonormal eigenvector basis~\cite{Pumplin:2001ct}. }
\end{center}
\end{figure}

With the covariance matrix approach, we can define
  $\chi^2=\frac{1}{N_{\rm dat}}\sum_{i,j}
  (d_i-\bar d_i){\rm cov}_{ij}(d_j-\bar d_j)$, $\bar d_i$ are data,
  $d_i$ theoretical predictions, ${N_{\rm dat}}$
  is the number of data points (note the inclusion of the factor
  $\frac{1}{N_{\rm dat}}$ in the definition) and ${\rm cov}_{ij}$ is
  the covariance matrix. Different groups may use somewhat different
  definitions of the covariance matrix, by including entirely or only
  partially correlated uncertainties. 
The best fit is the point in parameter
space at which $\chi^2$ is minimum, while PDF uncertainties are found
by diagonalizing the (Hessian) matrix of second
derivatives of the $\chi^2$ at the minimum (see
Fig.~\ref{fig:hessian}) and then determining the range of each
orthonormal Hessian eigenvector which corresponds to a prescribed
increase of the $\chi^2$ function with respect to the minimum.

In principle, the variation of the $\chi^2$ which corresponds to a
68\% confidence (one sigma) is $\Delta\chi^2=1$.  However, a larger variation
 $\Delta\chi^2=T^2$, with $T>1$ a suitable ``tolerance''
 parameter~\cite{Stump:2001gu,Pumplin:2001ct,Martin:2002aw} may
turn out to be necessary for more realistic error estimates for fits containing a wide variety of input processes/data, and in particular
in order for each
 individual experiment which enters the global fit to be consistent
 with the global best fit to one sigma (or some other desired
 confidence level 
such as 90\%).  Possible reasons why this is necessary could be related to
data inconsistencies or incompatibilities, 
underestimated experimental systematics, insufficiently flexible
parton parametrizations, theoretical uncertainties or approximation in
the PDF extraction. At present, HERAPDF and ABKM use $\Delta\chi^2=1$,
GJR uses $T\approx4.7$ at one sigma
(corresponding to $T\approx7.5 $ at 90\% c.l.), CTEQ6.6
uses $T=10$ at 90\% c.l. (corresponding to $T\approx6.1 $ to one
sigma) and MSTW08 uses a dynamical tolerance~\cite{Martin:2009iq}, i.e. a different value
of $T$ for each eigenvector, with values 
for one sigma
ranging from $T\approx 1$ to
$T\approx 6.5$ and most values being $2<T<5$. 

Within the NNPDF  method, PDFs are determined by first producing a
Monte Carlo sample of $N_{\rm rep}$ pseudo-data replicas. Each replica
contains a number of points equal to the number of original data
points.
 The sample is constructed
in such a way that, in the limit $N_{\rm rep}\to\infty$, 
the central value of the $i$-th
data point is equal to the mean over the $N_{\rm rep}$ values that the
$i$-th point takes in each replica,  the uncertainty of the
same point is  equal to the variance over the
replicas, and the correlations between any two original data
points is equal to their covariance over the replicas.
From each data replica, a PDF replica  is constructed by minimizing
a $\chi^2$ function. PDF
central values, uncertainties and correlations are then computed by
taking means, variances and covariances over this replica
sample. NNPDF uses a Monte Carlo method, with each PDF replica
obtained as the minimum $\chi^2$ which satisfies a cross-validation
criterion~\cite{Ball:2008by,Ball:2010de}, and is thus larger than the
absolute minimum of the $\chi^2$. This method has been used in all NNPDF sets
from NNPDF1.0 onwards. 

\subsubsection{Parton parametrization}
\label{parm}
Existing parton parametrizations differ  in the number of PDFs
which are independently parametrized and  in the functional  form and 
number of independent parameters used. They also differ in the choice
of individual linear combinations of PDFs which are parametrized.
In what concerns the functional form, the most common choice is that
each PDF at some reference scale $Q_0$ is parametrized as
\be
f_i(x,Q_0)=N x^{\alpha_i} (1-x)^{\beta_i} g_i(x)
\label{pdfparm}
\ee
where $g_i(x)$ is a function which tends to a constant both for $x\to1$
and $x\to0$, such as for instance $g_i(x)=1+ \epsilon_i \sqrt{x}+ D_i x+ E_i
x^2$ (HERAPDF). The fit parameters are $\alpha_i$, $\beta_i$ and the
parameters in $g_i$. Some of these parameters may be chosen to take a
fixed value (including zero).
The general form Eq.~(\ref{pdfparm})  is adopted in
all PDF sets which we discuss here except NNPDF, which instead lets
\be
f_i(x,Q_0)= c_i(x) NN_i(x)
\label{pdfparmnn}
\ee
where $NN_i(x)$ is a neural network, and $c_i(x)$ is is a
``preprocessing'' function. The fit parameters are the parameters
which determine the shape of the neural network (a 2-5-3-1
feed-forward neural network for NNPDF2.0). The preprocessing function
is not fitted, but rather chosen randomly in a space of functions of
the general form Eq.~(\ref{pdfparmnn}) within some acceptable range of
the parameters $\alpha_i$ and $\beta_i$, and with $g_i=1$.

The basis functions and number of parameters are the following.
\begin{itemize}
\item ABKM parametrizes the two lightest flavours and antiflavours,
  the total strangeness and the gluon (five independent PDFs)
  with
21 free parameters.
\item CTEQ6.6 and CT10  parametrize  the two lightest flavours and antiflavours
  the total strangeness and the gluon (six independent PDFs) with respectively 22
  and 26 free parameters.
\item GJR parametrizes the two lightest flavours and antiflavours and
  the gluon with 
20 free parameters (five independent PDFs); 
the strange distribution is assumed to be either
proportional to the light sea or to vanish at a low scale $Q_0<1$~GeV at which
PDFs become valence-like.
\item HERAPDF parametrizes the two lightest flavours, $\bar u$,
  the combination $\bar d+\bar s$ and the gluon with 10 free
  parameters (six independent PDFs), strangeness is assumed to be
  proportional to the $\bar d$ distribution; HERAPDF also studies the effect of
  varying the form of the 
  parametrization  and of
  and varying the relative size of the strange component and thus
  determine a model and parametrization uncertainty (see
  Sect.\ref{sec:herapdf} for more 
  details).   
\item  MSTW parametrizes the three lightest flavours and antiflavours and
  the gluon with 28 free parameters (seven independent PDFs) to find the 
best fit, but 8 are held fixed in determining uncertainty eigenvectors.
\item NNPDF parametrizes the three lightest flavours and antiflavours and
  the gluon with 259 free parameters (37 for each of the seven
  independent PDFs).
\end{itemize}

\subsection{PDF delivery and usage}
\label{delusage}

The way uncertainties should be determined for a given PDF set depends
on whether it is a Monte Carlo set (NNPDF) or a Hessian set (all other
sets). We now describe the procedure to be followed in each case. 

\subsubsection{Computation of  Hessian PDF uncertainties}

For Hessian PDF sets, both a central set and error sets are given. The
number of eigenvectors is equal to the number of free parameters. Thus, the number of error PDFs is equal to twice that. 
Each error
set corresponds to moving by the specified confidence level (one sigma
or 90\% c.l.) in the positive or negative direction of each independent orthonormal Hessian eigenvector. 

Consider a variable $X$; its value using the central PDF for an error set is given by $X_0$. $X_i^+$
is the value of that variable using the PDF corresponding to the ``$+$'' direction for the eigenvector $i$, and $X_i^-$ the
value for the variable using the PDF corresponding to the ``$-$'' direction. 

\begin{eqnarray}  
\Delta X^{+}_{\rm max}&=&\sqrt{\sum_{i=1}^N[max(X^{+}_{i}-X_{0},X^{-}_{i}-X_{0},0)]^{2}} \nonumber \\
\Delta X^{-}_{\rm max}&=&\sqrt{\sum_{i=1}^N[max(X_{0}-X^{+}_{i},X_{0}-X^{-}_{i},0)]^{2}}
\end{eqnarray}  

$\Delta X^{+}$ adds in quadrature the PDF error contributions that lead to an increase in the observable $X$, and
$\Delta X^{-}$ the PDF error contributions that lead to a decrease. The addition in quadrature is justified by the
eigenvectors forming an orthonormal basis. The sum is over all $N$ eigenvector directions. Ordinarily, one of $X^{+}_{i}-X_{0}$ and $X^{-}_{i}-X_{0}$ will be
positive and one will be negative, and thus it is
trivial as to which term is to be included in each quadratic sum. For the higher number (less well-determined) eigenvectors, however,  the ``$+$'' and ``$-$''eigenvector contributions may be in the same direction. In this case, only the more positive term will be included in the calculation of
$\Delta X^{+}$ and the more negative in the calculation of $\Delta X^{-}$~\cite{Nadolsky:2001yg}.
Thus, there may be less than $N$ non-zero terms for either
the ``$+$'' or ``$-$'' directions. A symmetric version of this is also used
by many groups, given by the equation below:

\begin{eqnarray}  
\Delta X&=&{1\over2}\sqrt{\sum_{i=1}^N[X^{+}_{i}-X^{-}_{i}]^{2}} \nonumber \\
\label{eq:symm}
\end{eqnarray}  

In most cases, the symmetric and asymmetric forms give very similar results. The extent to which the symmetric and asymmetric errors do not agree is an indication of the deviation of the $\chi^2$ distribution from a quadratic form. 
The lower number eigenvectors, corresponding to the best known
directions in eigenvector space, tend to have very symmetric errors,
while the higher number eigenvectors can have asymmetric errors. The
uncertainty for a particular observable then will (will not) tend to
have a quadratic form if it is most sensitive to lower number (higher
number) eigenvectors.  Deviations from a quadratic form are expected
to be greater for larger excursions, i.e. for 90\%c.l. limits than for
68\% c.l. limits.

The HERAPDF analysis also works with the Hessian matrix, defining experimental error PDFs 
in an orthonormal basis as described above. The symmetric formula Eq.~\ref{eq:symm} is most often used to calculate the 
experimental error bands on any variable, but it is possible to use the asymmetric formula as for MSTW and CTEQ. (For HERAPDF1.0 these errors are 
provided at $68\%$ c.l. in the LHAPDF file: HERAPDF10 EIG.LHgrid).

Other methods of calculating the PDF uncertainties independent of the Hessian method, such as the Lagrange Multiplier approach~\cite{Stump:2001gu}, are not
discussed here.

\subsubsection{Computation of   Monte Carlo PDF uncertainties}

For the NNPDF Monte Carlo set, a Monte Carlo sample of PDFs is given.
The expectation value of any observable 
$ \mathcal{F} [ \{  q \}]$ (for example a cross--section) which depends
on the PDFs is computed as an average over the ensemble of PDF
replicas, using
the following master formula:
\be
\label{masterave}
\la \mathcal{F} [ \{  q \}] \ra
= \frac{1}{N_{\rm rep}} \sum_{k=1}^{N_{\rm rep}}
\mathcal{F} [ \{  q^{(k)} \}],
\ee
where $N_{\rm rep}$ is the number of replicas of PDFs in the
Monte Carlo ensemble.
The associated uncertainty is found as the standard deviation of the
sample, according to the usual formula
\bea
\sigma_{\mathcal{F}} 
&=& \left( \frac{N_{\rm rep}}{N_{\rm rep}-1}   
\lp \la \mathcal{F} [ \{  q \}]^2\ra 
-   \la \mathcal{F} [ \{  q \}] \ra^2 
\rp \right)^{1/2}\nn\\
&=& \left( \frac{1}{N_{\rm rep}-1}
\sum_{k=1}^{N_{\rm rep}}   
\lp \mathcal{F} [ \{  q^{(k)} \}] 
-   \la \mathcal{F} [ \{  q \}] \ra\rp^2 
 \right)^{1/2}.
\label{mastersig}
\eea
These formulae may also be used for the determination of central values and
uncertainties of the parton distribution themselves, in which case the
functional $\mathcal{F}$ is identified with the parton distribution $q$ :  
$\mathcal{F}[ \{ q\}]\equiv q$.
Indeed, the central
value for PDFs themselves is given by 
 \be
\label{mcav}
q^{(0)} \equiv \la  q \ra = \frac{1}{N_{\rm rep}}
\sum_{k=1}^{N_{\rm rep}} q^{(k)} \ .
\ee

NNPDF provides both sets of $N_{\rm rep}=100$ and $N_{\rm rep}=1000$
replicas. The larger set ensures that statistical fluctuations are
suppressed so that even 
oddly-shaped probability distributions such as
non-gaussian or asymmetric ones are well reproduced, and more detailed
features of the probability distributions such as correlation
coefficients or uncertainties on uncertainties can be determined
accurately. However, for most common applications such as the
determination of the uncertainty on a cross section the smaller
replica set is adequate, and in fact central values can be determined
accurately using a yet smaller number of PDFs (typically $N_{\rm
  rep}\approx 10$), with the full set of $N_{\rm
  rep}\approx 100$ only needed for the reliable determination of
uncertainties.

NNPDF also provides a set 0 in the NNPDF20\_100.LHgrid LHAPDF
file, as in previous releases of the
NNPDF family, while replicas 1 to 100 correspond to
PDF sets 1 to 100 in the same file. This set 0 contains the average of
the PDFs, determined using Eq.~(\ref{mcav}): in other words, set 0
contains the central NNPDF prediction for each PDF. This central
prediction can be used to get a quick evaluation of a central
value. However, it should be noticed that for any
$\mathcal{F}[ \{  q \}]$ which 
depends nonlinearly on the PDFs, 
$\la \mathcal{F} [ \{  q \}] \ra 
\not= 
 \mathcal{F} [\{  q^{(0)}\}]$. This means that a cross section
 evaluated from the central set is not exactly equal to the central
 cross section (though it will be for example for deep-inelastic
 structure functions, which are linear in the PDFs). Hence,
use of the 0 set  is not 
recommended for precision applications, 
though in most cases it will provide a
good approximation.
Note that set $q^{(0)}$ should not be included when
computing an average with Eq.~(\ref{masterave}), because it is 
itself already an average. 

Equation~(\ref{mastersig}) provides
the 1--sigma PDF uncertainty on a general quantity which
depends on PDFs. However, 
an important advantage of the Monte Carlo method is that one does not
have to rely on a Gaussian assumption or on linear
error propagation. As a consequence, one may determine directly a
confidence level: e.g. a 68\% c.l.  for $\mathcal{F}[ \{  q \}]$ is simply
found by computing the $N_{\rm rep}$ values of $\mathcal{F}$ and
discarding the upper and lower 16\% values.
In a general non-gaussian case this 68\% c.l.
might be asymmetric and not equal to the variance (one--sigma
uncertainty). 
For the observables of the present
benchmark study the 1--sigma and 68\% c.l. PDF uncertainties
turn out to be very similar and thus only the former are given,
but this is not necessarily the case in
in general. For example, the one sigma error band on the NNPDF2.0 large
$x$ gluon and the small $x$ strangeness is much larger than the corresponding 68\% CL band, suggesting non-gaussian behavior of the probability distribution in these regions, in which PDFs are being extrapolated beyond the data region.

\clearpage

\section{PDF determinations - Theoretical uncertainties}
\label{sec:pdfdet2}

Theoretical uncertainties of PDFs
 determined in global fits reflect the approximations in the theory
 which is used in order to relate PDFs to measurable quantities.
The study of theoretical PDF uncertainties is currently less advanced
that that of experimental uncertainties, and only some theoretical
uncertainties have been explored. One might expect that the main
theoretical uncertainties in PDF determination
should be related to the treatment of the strong interaction: in
particular to the values of the QCD parameters, specifically the value
of the strong coupling  $\alpha_s$ and of the quark masses  $m_c$ and
$m_b$ and  uncertainties related to the truncation of the perturbative
expansion (commonly estimated through the variation of renormalization
and factorization scales). Further uncertainties are related to the
treatment of heavy quark thresholds, which are handled in various ways
by different groups (fixed flavour number vs. variable flavour number
schemes, and in the latter case different implementations of the
variable flavour number scheme), and to further approximations such as
the use of $K$-factor approximations.
Finally,  more uncertainties may be related to
weak interaction parameters (such as the $W$ mass) and to the
treatment of electroweak effects (such as QED PDF evolution~\cite{Martin:2004dh}
).

Of these
uncertainties, the only one which has been explored systematically by
the majority of the PDF groups is the $\alpha_s$ uncertainty. The way
$\alpha_s$ uncertainty can be determined using CTEQ, HERAPDF, MSTW,
and NNPDF will be discussed in detail below. HERAPDF also provides 
model and parametrization uncertainties which include 
the effect of varying $m_b$ and
$m_c$,
as well as the effect of varying the parton parametrization, as
will also be discussed below. Sets with varying quark masses and their implications have 
recently been made available by MSTW \cite{Martin:2010db}, the effects of varying $m_c$ and $m_b$ have been included by ABKM~\cite{ Alekhin:2009ni} and preliminary studies of the effect of $m_b$ and $m_c$ have also been presented by NNPDF~\cite{Rojo:2010gv}. Uncertainties related to factorization
and renormalization scale variation and to electroweak effects are so
far not available. For the benchmarking exercise of 
Sec.~\ref{sec:benchmarking}, results are given adopting
common values of electroweak parameters, 
and at least one common value of $\alpha_s$ (though
values for other values of $\alpha_s$ are also given), but no attempt
has yet been made to benchmark the other aspects mentioned above.

\subsection{The value of $\alpha_s$ and its uncertainty}

We thus turn to the only theoretical uncertainty which has been
studied systematically so far, namely the uncertainty on $\alpha_s$.
The choice of value of $\alpha_s$ is clearly important because it
is strongly correlated to PDFs, especially the gluon distribution (the
correlation of $\alpha_s$ with the gluon distribution using CTEQ, MSTW
and NNPDF  PDFs is studied in detail in Ref.~\cite{Demartin:2010er}). See also Ref.~\cite{ Alekhin:2009ni} for a discussion of this correlation in the ABKM PDFs. 
There are two separate issues related to the value of $\alpha_s$ in
PDF fits: first, the choice of $\alpha_s(m_Z)$ for which PDFs are made available, and second the choice of the preferred value of $\alpha_s$ to be
used when giving PDFs and their uncertainties. 
The two issues are related but independent, and for each
of the two issue two different basic philosophies 
may be adopted.

Concerning the range of available values
of $\alpha_s$:

\begin{itemize}
\item PDFs fits are performed for a number of different values of
  $\alpha_s$. Though a 
 PDF set corresponding to some
  reference value of $\alpha_s$ is given, the user is free to
  choose any of the given sets. This approach is adopted by CTEQ (0.118),
  HERAPDF (0.1176), MSTW (0.120) and NNPDF (0.119), where we have
  denoted in parenthesis the reference (NLO) value of $\alpha_s$ for each set.
\item $\alpha_s(m_Z)$  is treated as a fit parameters and PDFs
  are given only for the best--fit value. This approach is adopted
  by ABKM (0.1179) and GJR (0.1145), 
where in parenthesis the best-fit (NLO) value of $\alpha_s$ is given.
\end{itemize}

\begin{figure}
\begin{center}
\includegraphics[height=65mm,angle=0]{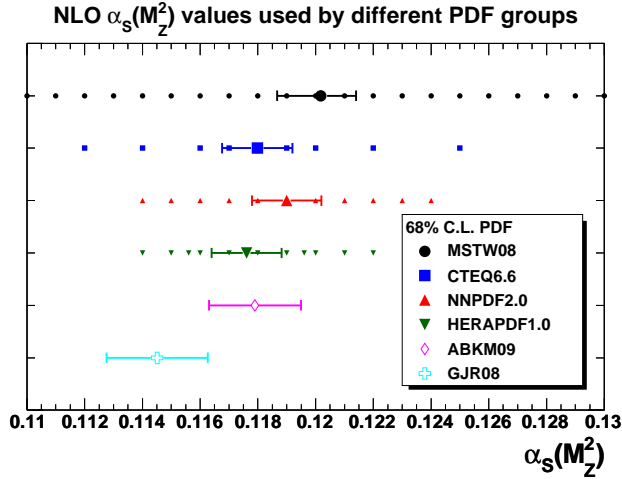}
\caption{\label{fig:alphas}  
Values of $\alpha_s(m_Z)$ for which fits are available. The 
default values and uncertainties used by each group are also shown.
Plot by G. Watt \cite{Watt}.}
\end{center}
\end{figure}

Concerning the preferred central value and the  
treatment of the $\alpha_s$ uncertainty:
\begin{itemize}
\item The value of $\alpha_s(m_Z)$ is taken as an external parameter,
  along with other parameters of the fit such as heavy quark masses or
  electroweak parameter. This approach is adopted by CTEQ,
  HERAPDF1.0  and NNPDF. In this case, there is no apriori central value of $\alpha_s(m_Z)$ and the uncertainty on
  $\alpha_s(m_Z)$ is treated by repeating the PDF determination as
  $\alpha_s$ is varied in a suitable range. Though a 
 range
  of variation is usually chosen by the groups, any other range may be
  chosen by the user.
\item The value of $\alpha_s(m_Z)$ is treated as a fit parameter, and
  it is determined along with the PDFs. This  approach is adopted by
  MSTW, ABKM and GJR08. In the last two cases, the uncertainty on $\alpha_s$ is part of the Hessian matrix of the fit. The MSTW approach is explained below.
\end{itemize}
As a cross-check,CTEQ~\cite{Lai:2010nw} has also used the 
world average value of $\alpha_s(m_Z)$ 
as an additional input to the global fit.

The values of $\alpha_s(m_Z)$ for which fits are available, as well as
the default values and uncertainties used by each group are 
summarized in Fig.~\ref{fig:alphas}~\footnote{There is implicitly an additional uncertainty due to scale variation.See for example Ref.~\cite{Blumlein:1996gv}.}.  
The most recent world average value of $\alpha_s(m_Z)$ is
 $\alpha_s = 0.1184\pm0.0007$~\cite{Bethke:2009jm}~\footnote{We note that the values used in the average are from extractions 
at different orders in the perturbative expansion.}. However, a more
conservative estimate of the uncertainty on $\alpha_s$ was felt to be
appropriate  for the benchmarking exercise summarized in this note,
for which we have taken $\Delta\alpha_s=\pm0.002$ at
90\%c.l. (corresponding to 0.0012 at one sigma). This uncertainty 
has been used for the CTEQ, NNPDF and HERAPDF studies. For MSTW, ABKM and 
GJR the preferred $\alpha_s$ uncertainty for each group is used, though for MSTW
in particular this is close to 0.0012 at one sigma. 
It may not be unreasonable to argue that a yet 
larger uncertainty may be appropriate. 


When comparing results obtained using different PDF sets it should be
borne in mind that if different values of $\alpha_s$ are used,
cross section predictions change both because of the dependence of the
cross section on the value of $\alpha_s$ (which for some processes
such as top production or Higgs production in gluon-gluon fusion may
be quite strong), and because of the dependence of the PDFs themselves on the
value of $\alpha_s$. Differences due to the PDFs alone can be isolated
only when performing comparisons at a common value of $\alpha_s$.


\subsection{Computation of PDF+$\alpha_s$ uncertainties}
\label{pdfas}

Within the quadratic approximation to the dependence of $\chi^2$ on
parameters (i.e. linear error propagation), it turns out that even if
PDF  uncertainty and the $\alpha_s(m_Z)$ uncertainty are correlated,
the total one-sigma 
combined PDF+$\alpha_s$ uncertainty including this
correlation can be simply found without approximation by computing the
one sigma PDF uncertainty with $\alpha_s$ fixed at its central value  
and the one-sigma  $\alpha_s$ uncertainty with the PDFs fixed their
central value, and adding results in quadrature~\cite{Lai:2010nw}, and
similarly for any other desired confidence level. 

For example, if
$\Delta X_{PDF}$ is the PDF uncertainty for a cross section $X$ and
$\Delta X_{\alpha_s(m_Z)}$ is the $\alpha_s$ uncertainty, the combined
uncertainty $\Delta X$ is  
\be  
\Delta X=\sqrt{\Delta X_{PDF}^2 + \Delta X_{\alpha_s(m_Z)}^2}\label{pdfasunc}
\ee

Other treatments can be used when deviations
from the quadratic approximation are possible. 
Indeed,for MSTW because of the use of
dynamical tolerance linear error propagation does not necessarily
apply. For NNPDF, because of the use of a Monte Carlo method linear
error propagation is not assumed: in practice, addition in quadrature
turns out to be a very good approximation, but an exact treatment is
computationally simpler.
We now describe in detail the procedure for the computation of
$\alpha_s$ and PDF uncertainties (and for HERAPDF also of model and
parametrization 
uncertainties) for various parton sets.

\subsubsection{CTEQ  - Combined PDF and $\alpha_s$ uncertainties}

CTEQ takes $\alpha_s^0(m_Z)=0.118$ 
as an external input parameter and provides
the CTEQ6.6alphas \cite{Lai:2010nw} (or the CT10alpha~\cite{Lai:2010vv}) series
which contains 4 sets extracted using
$\alpha_s(m_Z)=0.116,~0.117,~0.119,~0.120$;
The uncertainty associated with \as~ can be evaluated by computing any
given observable with $\alpha_s=0.118\pm\delta^{(68)}$ in the partonic
cross-section and with the PDF sets that have been extracted
with these values of \as.
The differences
\be
\Delta_+^{\alpha_s}={\cal F}(\alpha_s^0+\delta^{(68)}\alpha_s)-{\cal F}(\alpha_s^0),
\quad\quad
\Delta_-^{\alpha_s}={\cal F}(\alpha_s^0-\delta^{(68)}\alpha_s)-{\cal F}(\alpha_s^0)
\label{ascteq}
\ee
are the \as~ uncertainties according to CTEQ.
In \cite{Lai:2010nw} it has been demonstrated that, in the Hessian approach,
the combination in quadrature of PDF and \as uncertainties is
correct within the quadratic approximation. In the studies in Ref.~\cite{Lai:2010nw}, CTEQ did not find appreciable deviations from the quadratic approximation, and thus the procedure described below will be accurate for the cross sections considered here. 

Therefore,
for CTEQ6.6 the combined PDF+\as uncertainty is
given by
\bea
\Delta^{PDF+\alpha_s}_+&=&
\sqrt{\left(\Delta_+^{\alpha_s} \right)^2 +  
      \left((\Delta F_{PDF}^{\alpha_s^0})_+\right)^2}
\label{pdfascteq}
\\
\Delta^{PDF+\alpha_s}_-&=&
\sqrt{\left(\Delta_-^{\alpha_s} \right)^2 +  
      \left((\Delta F_{PDF}^{\alpha_s^0})_-\right)^2}
\nonumber
\eea

\subsubsection{MSTW  - Combined PDF and $\alpha_s$ uncertainties}

MSTW fits \as~ together with the PDFs
and obtains $\alpha_s^0(NLO)=0.1202^{+0.0012}_{-0.0015}$ and 
$\alpha_s^0(NNLO)=0.1171\pm0.0014$.
Any correlation between the PDF and the \as~ uncertainties is
taken into account with the following recipe
\cite{Martin:2009bu}.
Beside the best-fit sets of PDFs, which correspond to
$\alpha_s^0(NLO,NNLO)$, 
four more sets,both at NLO and at NNLO, of PDFs are provided.
The latter are extracted setting as input 
$\alpha_s=\alpha_s^0\pm 0.5 \sigma_{\alpha_s}, \alpha_s^0\pm
\sigma_{\alpha_s}$, where $\sigma_{\alpha_s}$ is the standard
deviation indicated here above.
Each of these extra sets contains the full parametrization to describe
the PDF uncertainty. 
Comparing the results of the five sets, 
the combined PDF+\as uncertainty is defined as:
\bea
\Delta^{PDF+\alpha_s}_+&=&
\max_{\alpha_s}\left\{ F^{\alpha_s}(S_0)+(\Delta F^{\alpha_s}_{PDF})_+ \right\}
- F^{\alpha_s^0}(S_0)\label{pdfasmstw}\\
\Delta^{PDF+\alpha_s}_-&=&
F^{\alpha_s^0}(S_0)\nonumber
-
\min_{\alpha_s}\left\{ F^{\alpha_s}(S_0)-(\Delta F^{\alpha_s}_{PDF})_- \right\}
\eea
where $\max,\min$ run over the five values of $\alpha_s$ under study,
and the corresponding PDF uncertainties are used.

The central and $\alpha_s=\alpha_s^0\pm 0.5 \sigma_{\alpha_s}, \alpha_s^0\pm
\sigma_{\alpha_s}$, where $\sigma_{\alpha_s}$ sets are all obtained using the 
dynamical tolerance prescription for PDF uncertainty which determines the 
uncertainty when the quality of the fit to any one data set (relative to the best 
fit for the preferred value of $\alpha_s(M_Z)$) becomes sufficiently poor. Naively 
one might expect that the PDF uncertainty for the $\alpha_s^0\pm
\sigma_{\alpha_s}$ might then be zero since one is by definition already at the 
limit of allowed fit quality for one data set. If this were the case the 
procedure of adding PDF and $\alpha_S$ uncertainties would be a very good 
approximation. However, in practice there is freedom to move the PDFs in particular
directions without the data set at its limit of fit quality becoming worse fit, and 
some variations can be quite large before any data set becomes sufficiently badly 
fit for the criterion for uncertainty to be met. This can led to significantly
larger PDF $+ \alpha_s$ uncertainties than the simple quadratic prescription. 
In particular, since there is a tendency for the best fit to have a too low value
of $d F_2/d\ln Q^2$ at low $x$, at higher $\alpha_s$ value the small-$x$ gluon 
has freedom to increase without spoiling the fit, and the PDF $+\alpha_S$ uncertainty 
is large in the upwards direction for Higgs production.

\subsubsection{HERAPDF - $\alpha_s$,  model and parametrization uncertainties}
\label{sec:herapdf}

HERAPDF provides not only $\alpha_s$ uncertainties, but also model and
parametrization uncertainties. Note that at least in part
parametrization uncertainty will be accounted for by other groups by
the use of a significantly larger number of initial parameters,
the use of a large tolerance (CTEQ, MSTW) or by a more general
parametrization (NNPDF), as discussed in Sect.~\ref{parm}. However,
model uncertainties related to heavy quark masses are not determined
by other groups.

The model errors come from variation of the choices of: charm 
mass ($m_c=1.35 \to 1.65$GeV); beauty mass ($m_b=4.3 \to 5.0$~GeV); minimum $Q^2$ of data 
used in the fit ($Q^2_{min}=2.5 \to 5.0$~GeV$^2$); fraction of strange sea in total d-type 
sea ($f_s=0.23 \to 0.38$ at the starting scale). The model errors are calculated by taking 
the difference between the central fit and the model variation and adding them in 
quadrature, separately for positive and negative deviations. (For HERAPDF1.0 the model
variations are provided as members 1 to 8 of the LHAPDF file: HERAPDF10 VAR.LHgrid).

The parametrization errors come from: variation of the starting scale $Q^2_0=1.5 \to 2.5$
GeV$^2$; 
 variations of the basic 10 parameter fit to 11 parameter fits in which an extra parameter 
is allowed to be free for each fitted parton distribution. In practice only three of 
these extra parameter variations have significantly different PDF shapes from the central fit. The 
parametrization errors are calculated by storing the difference between the 
parametrization variant and the central fit and constructing an envelope representing 
the maximal deviation at each $x$ value. (For HERAPDF1.0 the parametrization variations are 
provided as members 9 to 13 of the LHAPDF file: HERAPDF10 VAR.LHgrid).

HERAPDF also provide an estimate of the additional error due to the uncertainty on $\alpha_s(M_Z)$.
 Fits are made with the central value, $\alpha_s(M_Z)=0.1176$, varied by $\pm 0.002$. 
The $90\%$ c.l. $\alpha_s$ error on any variable should be calculated by adding in quadrature the 
difference between its value as calculated using the central fit and its value using these two 
alternative $\alpha_s$ values; $68\%$ c.l. values may be obtained by scaling the result down by 1.645. (For HERAPDF1.0 these $\alpha_s$ variations are 
provided as members 9,10,11 of the LHAPDF file: HERAPDF10 ALPHAS.LHgrid for $\alpha_s(M_Z)=0.1156,0.1176,0.1196$, respectively). 
Additionally members 1 to 8 provide PDFs for values of $\alpha_s(M_Z)$ ranging
from 0.114 to 0.122). The total PDF + $\alpha_s$ uncertainty for HERAPDF should be constructed by adding in quadrature 
experimental, model, parametrization and $\alpha_s$ uncertainties. 

\subsubsection{NNPDF - Combined PDF and $\alpha_s$ uncertainties}
\label{sec:NNPDFdetails}

For the NNPDF2.0 family, PDF sets obtained with values of $\alpha_s(m_Z)$
in the range from 0.114 to 0.124 in steps of $\Delta\alpha_s=0.001$ are
available in LHAPDF. Each of these sets is denoted by
NNPDF20\_as\_0114\_100.LHgrid, NNPDF20\_as\_0115\_100.LHgrid, ... and
has the same structure as the central NNPDF20\_100.LHgrid set: PDF set
number 0 is the average PDF set, as discussed above
 \be
\label{mcavalphas}
q^{(0)}_{\alpha_s} \equiv \la  q_{\alpha_s} \ra = \frac{1}{N_{\rm rep}}
\sum_{k=1}^{N_{\rm rep}} q^{(k)}_{\alpha_s} \ .
\ee 
for the different values of $\alpha_s$, while sets from 1 to 100
are the 100 PDF replicas corresponding to this particular value of
$\alpha_s$. Note that in general not only the PDF
central values but also the PDF uncertainties will depend 
on $\alpha_s$.

The methodology used within the NNPDF approach to combine
PDF and $\alpha_s$ uncertainties is
 discussed in Ref.~\cite{Demartin:2010er,LH}.
One possibility is to add in quadrature the
PDF and $\alpha_s$ uncertainties,  using PDFs
obtained from different values of  $\alpha_s$, which as discussed
above is correct in the quadratic approximation. However use of the
exact correlated Monte Carlo formula turns out to be actually simpler,
as we now show.

If the sum in quadrature is adopted, 
for a generic cross section which depends on the PDFs and the
strong coupling  $\sigma\lp {\rm PDF},\alpha_s\rp$, we have
\be
\label{LH_NNPDF_eq:deltaalphas2}
\lp \delta \sigma \rp_{\alpha_s}^{\pm}
=\sigma\lp {\rm PDF}^{(\pm)},\alpha_s^{(0)}\pm \delta_{\alpha_s} \rp - \sigma\lp {\rm PDF}^{(0)},\alpha_s^{(0)} \rp  \ ,
\ee
where {\rm PDF}$^{(\pm)}$ stands schematically for the PDFs obtained
when $\alpha_s$ is varied within its 1--sigma range, $\alpha_s^{(0)}
\pm \delta_{\alpha_s}$.
The PDF+$\alpha_s$ uncertainty is
\be
\label{LH_NNPDF_eq:deltasigma}
\lp \delta \sigma \rp_{{\rm PDF}+\alpha_s}^{\pm} =\sqrt{
\lc \lp \delta \sigma \rp_{\alpha_s}^{\pm}\rc^2 +
\lc \lp \delta \sigma \rp_{\rm PDF}^{\pm}\rc^2 } \ .
\ee
with $\lp \delta \sigma \rp_{\rm PDF}^{\pm}$ the PDF uncertainty on the
observable $\sigma$ computed from the set with the central
value of $\alpha_s$.

The exact Monte Carlo expression instead is found
noting that
the average over Monte Carlo replicas  
of a general quantity which depends on both
$\alpha_s$ and the PDFs, $\mathcal{F}\lp  {\rm PDF},\alpha_s\rp$
is 
\be
\label{LH_NNPDF_eq:avrep}
\la \mathcal{F}\ra_{\rep} =\frac{1}{N_{\rep}}\sum_{j=1}^{N_{\alpha}}
\sum_{k_j=1}^{N_{\rm rep}^{\alpha_s^{(j)}}} 
\mathcal{F}\lp  {\rm PDF}^{(k_j,j)},\alpha_s^{(j)}\rp \ ,
\ee
where ${\rm PDF}^{(k_j,j)}$ stands for the replica $k_j$ of the
PDF fit obtained using $\alpha_s^{(j)}$ as the value of the
strong coupling; $N_{\rm rep}$ is
the total number of PDF replicas 
\be
N_{\rm rep} = \sum_{j=1}^{N_{\alpha_s}}N^{\alpha_s^{(j)}}_{\rm rep} \ ;
\ee
and  $N^{\alpha_s^{(j)}}_{\rm rep}$  is the number of PDF replicas
for each value $\alpha_s^{(j)}$ of $\alpha_s$. If we assume that 
$\alpha_s$ is gaussianly distributed about its central value with
width equal to the stated uncertainty,
the number of replicas
for each different value of $\alpha_s$ 
is 
\be
N^{\alpha_s^{(j)}}_{\rm rep}\propto \exp\lp 
-\frac{\lp \alpha_s^{(j)}- \alpha_s^{(0)}\rp^2}{
2 \delta_{\alpha_s}^2}\rp \ .
\ee
with $\alpha_s^{(0)}$ and $\delta_{\alpha_s}$ the assumed
central value and 1--sigma uncertainty of $\alpha_s(m_Z)$. Clearly
with a Monte Carlo method a different probability distribution of
$\alpha_s$ values could also be assumed. For example, if we assume
$\alpha_s(M_z)=0.119\pm0.0012$ and we take nine distinct values
$\alpha_s^{(j)}=0.115, 0.116, 0.117, 
0.118,$
$~0.119, 0.120, 0.123$,
assuming 100 replicas for the central value ($\alpha_s=0.119$) we get
$N^{\alpha_s^{(j)}}_{\rm rep}=0,4,25,71,$
$100,71,25,4,0$.

The  combined PDF+$\alpha_s$ uncertainty is then simply found by using
Eq.~(\ref{mastersig}) with averages computed using
Eq.~(\ref{LH_NNPDF_eq:avrep}).
The difference between Eq.~(\ref{LH_NNPDF_eq:avrep}) and
Eq.~(\ref{LH_NNPDF_eq:deltasigma}) measures deviations from linear
error propagation.
The NNPDF benchmark results presented below are obtained using
Eq.~(\ref{LH_NNPDF_eq:avrep}) with
$\alpha_s\lp m_Z\rp=0.1190\pm 0.0012$ at one sigma. No significant
deviations from linear error propagation were observed.

It is interesting to observe that the same method
can be used to determine the combined uncertainty of PDFs and
other physical parameters, such as heavy quark masses.


\section{PDF correlations}
\label{sec:corr}

The uncertainty analysis may be extended to define a \emph{correlation}
between the uncertainties of two variables, say $X(\vec{a})$ and
$Y(\vec{a}).$  As for the case of PDFs, the physical concept of PDF
correlations can be determined both from PDF determinations based
on the Hessian approach and on the Monte Carlo approach.

\subsection{PDF correlations in the Hessian approach}

Consider the projection of the tolerance hypersphere
onto a circle of radius 1 in the plane of the gradients $\vec{\nabla}X$
and $\vec{\nabla}Y$ in the parton parameter space \cite{Pumplin:2001ct,Nadolsky:2001yg}.
The circle maps onto an ellipse in the $XY$ plane. This {}``tolerance
ellipse'' is described by Lissajous-style parametric equations,\begin{eqnarray}
X & = & X_{0}+\Delta X\cos\theta,\label{ellipse1}\\
Y & = & Y_{0}+\Delta Y\cos(\theta+\varphi),\label{ellipse2}
\end{eqnarray}
 where the parameter $\theta$ varies between 0 and $2\pi$, $X_{0}\equiv X(\vec{a}_{0}),$
and $Y_{0}\equiv Y(\vec{a}_{0})$. $\Delta X$ and $\Delta Y$ are
the maximal variations $\delta X\equiv X-X_{0}$ and $\delta Y\equiv Y-Y_{0}$
evaluated according to the $Master$ Equation,
and $\varphi$ is the
angle between $\vec{\nabla}X$ and $\vec{\nabla}Y$ in the $\{ a_{i}\}$
space, with\begin{equation}
\cos\varphi=\frac{\vec{\nabla}X\cdot\vec{\nabla}Y}{\Delta X\Delta Y}=\frac{1}{4\Delta X\,\Delta Y}\sum_{i=1}^{N}\left(X_{i}^{(+)}-X_{i}^{(-)}\right)\left(Y_{i}^{(+)}-Y_{i}^{(-)}\right).\label{cosphi}\end{equation}

\begin{figure}
\begin{center}
\includegraphics[height=45mm,angle=0]{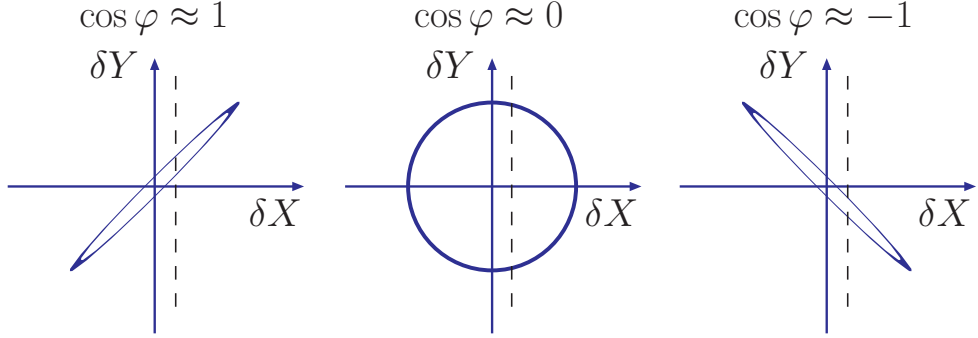}
\caption{Correlations ellipses  for a strong correlation (left), no correlation (center) and a strong anti-correlation(right)~\cite{Nadolsky:2008zw}. 
\label{fig:CorrelationEllipsePhi}}
\end{center}
\end{figure}

The quantity $\cos\varphi$ characterizes whether the PDF degrees
of freedom of $X$ and $Y$ are correlated ($\cos\varphi\approx1$),
anti-correlated ($\cos\varphi\approx-1$), or uncorrelated ($\cos\varphi\approx0$).
If units for $X$ and $Y$ are rescaled so that $\Delta X=\Delta Y$
(e.g., $\Delta X=\Delta Y=1$), the semimajor axis of the tolerance
ellipse is directed at an angle $\pi/4$ (or $3\pi/4)$ with respect
to the $\Delta X$ axis for $\cos\varphi>0$ (or $\cos\varphi<0$).
In these units, the ellipse reduces to a line for $\cos\varphi=\pm1$
and becomes a circle for $\cos\varphi=0$, as illustrated by Fig.~\ref{fig:CorrelationEllipsePhi}.
These properties can be found by diagonalizing the equation for the
correlation ellipse.
Its semiminor and semimajor axes (normalized to $\Delta X=\Delta Y$)
are\begin{eqnarray}
\{ a_{minor},a_{major}\} & = & \frac{\sin\varphi}{\sqrt{1\pm\cos\varphi}}.\label{ellipse4}\end{eqnarray}
 The eccentricity $\epsilon\equiv\sqrt{1-(a_{minor}/a_{major})^{2}}$
is therefore approximately equal to $\sqrt{\left|\cos\varphi\right|}$
as $\left|\cos\varphi\right|\rightarrow1$. 

\begin{equation}
\left(\frac{\delta X}{\Delta X}\right)^{2}+\left(\frac{\delta Y}{\Delta Y}\right)^{2}-2\left(\frac{\delta X}{\Delta X}\right)\left(\frac{\delta Y}{\Delta Y}\right)\cos\varphi=\sin^{2}\varphi.
\label{ellipse3}
\end{equation}

A magnitude of $|\cos\varphi|$ close to unity suggests that a precise
measurement of $X$ (constraining $\delta X$ to be along the dashed
line in Fig.~\ref{fig:CorrelationEllipsePhi}) is likely to constrain
tangibly the uncertainty $\delta Y$ in $Y$, as the value of $Y$
shall lie within the needle-shaped error ellipse. Conversely, $\cos\varphi\approx0$
implies that the measurement of $X$ is not likely to constrain $\delta Y$
strongly.%
\footnote{The allowed range of $\delta Y/\Delta Y$ for a given $\delta\equiv\delta X/\Delta X$
is $r_{Y}^{(-)}\leq\delta Y/\Delta Y\leq r_{Y}^{(+)},$ where $r_{Y}^{(\pm)}\equiv\delta\cos\varphi\pm\sqrt{1-\delta^{2}}\sin\varphi.$%
}

The values of $\Delta X,$ $\Delta Y,$ and $\cos\varphi$ are also
sufficient to estimate the PDF uncertainty of any function $f(X,Y)$
of $X$ and $Y$ by relating the gradient of $f(X,Y)$ to $\partial_{X}f\equiv\partial f/\partial X$
and $\partial_{Y}f\equiv\partial f/\partial Y$ via the chain rule:\begin{equation}
\Delta f=\left|\vec{\nabla}f\right|=\sqrt{\left(\Delta X\ \partial_{X}f\ \right)^{2}+2\Delta X\ \Delta Y\ \cos\varphi\ \partial_{X}f\ \partial_{Y}f+\left(\Delta Y\ \partial_{Y}f\right)^{2}}.\label{df}\end{equation}
Of particular interest is the case of a rational function $f(X,Y)=X^{m}/Y^{n},$
pertinent to computations of various cross section ratios, cross section
asymmetries, and statistical significance for finding signal events
over background processes \cite{Nadolsky:2001yg}. For rational functions
Eq.~(\ref{df}) takes the form\begin{equation}
\frac{\Delta f}{f_{0}}=\sqrt{\left(m\frac{\Delta X}{X_{0}}\ \right)^{2}-2mn\frac{\Delta X}{X_{0}}\ \frac{\Delta Y}{Y_{0}}\ \cos\varphi\ +\left(n\frac{\Delta Y\ }{Y_{0}}\right)^{2}}.\label{dfrat}\end{equation}
For example, consider a simple ratio, $f=X/Y$. Then $\Delta f/f_{0}$
is suppressed ($\Delta f/f_{0}\approx\left|\Delta X/X_{0}-\Delta Y/Y_{0}\right|$)
if $X$ and $Y$ are strongly correlated, and it is enhanced ($\Delta f/f_{0}\approx\Delta X/X_{0}+\Delta Y/Y_{0}$)
if $X$ and $Y$ are strongly anticorrelated.

As would be true for any estimate provided by the Hessian method,
the correlation angle is inherently approximate. Eq.~(\ref{cosphi})
is derived under a number of simplifying assumptions, notably in the
quadratic approximation for the $\chi^{2}$ function within the tolerance
hypersphere, and by using a symmetric finite-difference formula 
for $\{\partial_{i}X\}$ that may fail if $X$ is not monotonic. With
these limitations in mind, we find the correlation angle to be a convenient
measure of interdependence between quantities of diverse nature, such
as physical cross sections and parton distributions themselves. For example, in Section~\ref{sec:cteq66}, the correlations for the benchmark cross sections are given with respect to that for $Z$ production. As expected, the $W^+$ and $W^-$ cross sections are very correlated with that for the $Z$, while the Higgs cross sections are uncorrelated ($m_{Higgs}$=120 GeV) or anti-correlated ($m_{Higgs}$=240 GeV). Thus, the PDF uncertainty for the ratio of the cross section for a 240 GeV Higgs boson to that of the cross section for $Z$ boson production is larger than the PDF uncertainty for Higgs boson production by itself. 

A simple $C$ code (corr.C) is available from the PDF4LHC website that calculates the correlation cosine between any two observables given two text files that present the cross sections for each observable as a function of the error PDFs.

\subsection{PDF correlations in the Monte Carlo approach}

General correlations between PDFs and physical observables can be
computed within the Monte Carlo approach used by NNPDF using standard
textbook methods. To illustrate this point, let us compute the
the correlation coefficient 
$\rho[A,B]$ for two observables $A$ and $B$  which depend on PDFs
(or are PDFs themselves). This correlation coefficient in the
Monte Carlo approach is given by
\begin{equation}
  \label{eq:correlation}
  \rho[A,B]=\frac{N_{rep}}{(N_{rep}-1)}\frac{\langle A B\rangle_{\mathrm{rep}}
    - \langle A\rangle_{\mathrm{rep}}\langle B\rangle_{\mathrm{rep}} }
  {\sigma_A\sigma_B}
\end{equation}
where the averages are taken over ensemble of the $N_{\mathrm{rep}}$ values 
of the observables computed with the different replicas in the NNPDF2.0 set, 
and $\sigma_{A,B}$ are the standard deviations of the ensembles.
The quantity $\rho$ characterizes whether two observables (or PDFs) 
are correlated 
($\rho \approx 1$), anti-correlated ($\rho \approx -1$) or uncorrelated 
($\rho\approx 0$).

This correlation can be generalized to other cases, for example
to compute the correlation between PDFs and the value
of the strong coupling $\alpha_s(m_Z)$, as studied in
Ref.~\cite{Demartin:2010er,LH}, for any given
values of $x$ and $Q^2$.
For example, the correlation between the strong coupling and the
gluon at $x$ and $Q^2$ (or in general any other PDF) is defined as the
usual correlation between two probability distributions, namely
\be
\label{LH_NNPDF_eq:gcorr}
\rho \lc  \alpha_s\lp M_Z^2\rp,g\lp x,Q^2\rp\rc=
\frac{N_{rep}}{(N_{rep}-1)}
\frac{\la \alpha_s\lp M_Z^2\rp g\lp x,Q^2\rp \ra_{\rep}-
\la \alpha_s\lp M_Z^2\rp\ra_{\rep}\la g\lp x,Q^2\rp \ra_{\rep}
}{\sigma_{\alpha_s\lp M_Z^2\rp}\sigma_{g\lp x,Q^2\rp}} \ ,
\ee
where averages over replicas include PDF sets with varying
$\alpha_s$ in the sense of Eq.~(\ref{LH_NNPDF_eq:avrep}).
Note that the computation of this
correlation takes into account not only
the central gluons of the fits with different $\alpha_s$ but
also the corresponding uncertainties in each case.

\clearpage

\section{The PDF4LHC benchmarks}
\label{sec:benchmarking}

A benchmarking exercise was carried out to which all PDF groups were
invited to participate. This exercise considered only the-then most up to date published versions/most commonly used of NLO PDFs from  6 groups: ABKM09~\cite{Alekhin:2009ni},~\cite{Alekhin:2010iu}, CTEQ6.6~\cite{Nadolsky:2008zw}, GJR08~\cite{Gluck:2008gs},
HERAPDF1.0~\cite{herapdf10}, MSTW08~\cite{Martin:2009iq}, NNPDF2.0~\cite{Ball:2010de}. The benchmark cross sections were evaluated at NLO at both 7 and 14 TeV. We report here primarily on the 7 TeV results.   

All of the benchmark processes were to be calculated with the
following settings:

\begin{enumerate}

\item at NLO in the $\overline{MS}$ scheme
\item all calculation done in a the 5-flavor quark ZM-VFNS scheme, though 
each group uses a different treatment of heavy quarks
\item at a center-of-mass energy of 7 TeV 
\item for the central value predictions, and for $\pm68\%$ and $\pm90\%$ c.l. PDF uncertainties
\item with and without the $\alpha_s$ uncertainties, with the prescription for combining the PDF and $\alpha_s$ errors to be specified
\item repeating the calculation with a central value of $\alpha_s(m_Z)$ of 0.119.

\end{enumerate}

To provide some standardization, a gzipped version of MCFM5.7~\cite{mcfm} was prepared by John Campbell, using the specified parameters and exact input files for each process. It was allowable for other codes to be used, but they had to be checked against the MCFM output values. 

The processes included in the benchmarking exercise are given below. 

\begin{enumerate}

\item $W^+,W^-$ and $Z$ cross sections and rapidity distributions including the cross section ratios $W^+/W^-$ and ($W^++W^-)/Z$ and the $W$ asymmetry as a function of rapidity ([$W^+(y)-W^-(y)]/[W^+(y)+W^-(y)$]).

\vspace{2mm}

The following specifications were made for the $W$ and $Z$ cross sections:

\vspace{2mm}

\begin{enumerate}

\item $m_Z$=91.188 GeV

\item $m_W$=80.398 GeV

\item zero width approximation used

\item $G_F$=0.116637 X $10^{-5} GeV^{-2}$

\item $sin^2\theta_W$ = 0.2227

\item other EW couplings derived using tree level relations

\item BR($Z\rightarrow ll$) = 0.03366

\item BR($W\rightarrow l\nu$) = 0.1080

\item CKM mixing parameters from Eq. 11.27 of the PDG2009 CKM review

\item scales: $\mu_R=\mu_F$ = $m_Z$ or $m_W$

\end{enumerate}

\vspace{2mm}

\item $gg\rightarrow Higgs$ total cross sections at NLO in the Standard Model

The following specifications were made for the Higgs cross section. 

\vspace{2mm}

\begin{enumerate}

\item $m_H$ = 120, 180 and 240 GeV

\item zero Higgs width approximation, no branching ratios taken into account

\item top loop only, with $m_{top}$ = 171.3 GeV in $\sigma_o$

\item scales: $\mu_R=\mu_F=m_{Higgs}$

\end{enumerate}

\vspace{2mm}

\item $t\bar{t}$ cross section at NLO

\vspace{2mm}

\begin{enumerate}

\item $m_{top}$ = 171.3 GeV

\item zero top width approximation, no branching ratios

\item scales: $\mu_R=\mu_F=m_{top}$

\end{enumerate}
\end{enumerate}

\vspace{2mm}


The cross sections chosen are all important cross sections at the LHC, for standard model benchmarking for the case of the $W,Z$ and top cross sections
and discovery potential for the case of the Higgs cross sections. Both $q\bar{q}$ and $gg$ initial states are involved. The NLO $W$ and $Z$ cross sections have a small dependence on the value of $\alpha_s(m_Z)$, while the dependence is sizeable for both $t\bar{t}$ and Higgs production. 



















\subsection{Comparison between benchmark predictions}

Now we turn to compare the results of the various PDF sets for the
LHC observables with the common benchmark settings discussed above.
To perform a more meaningful comparison, 
it is useful to first introduce the idea of  differential parton-parton luminosities. Such luminosities, when multiplied by the dimensionless 
cross section $\hat{s}\hat{\sigma}$ for a given process, provide a useful estimate  of 
the size of an event cross section at the LHC. 
Below we define the differential parton-parton luminosity
$dL_{ij}/d\hat{s}$:

\begin{equation}
\frac{d L_{ij}}{d\hat{s}\,dy} = 
\frac{1}{s} \, \frac{1}{1+\delta_{ij}} \, 
[f_i(x_1,\mu) f_j(x_2,\mu) + (1\leftrightarrow 2)] \; .
\label{eq1}
\end{equation}
The prefactor with the Kronecker delta avoids double-counting in case the
partons are identical.  The generic parton-model formula 
\begin{equation}
\sigma = \sum_{i,j} \int_0^1 dx_1 \, dx_2 \, 
f_i(x_1,\mu) \, f_j(x_2,\mu) \, \hat{\sigma}_{ij}
\end{equation}
can then be written as 
\begin{equation}
\sigma = \sum_{i,j} \int \left(\frac{d\hat{s}}{\hat{s}} \right) 
\, \left(\frac{d L_{ij}}{d\hat{s}}\right) \, 
\left(\hat{s} \,\hat{\sigma}_{ij} \right) \; .
\label{eq:xseclum}
\end{equation}

\begin{figure}
\begin{center}
\includegraphics[width=0.48\textwidth]{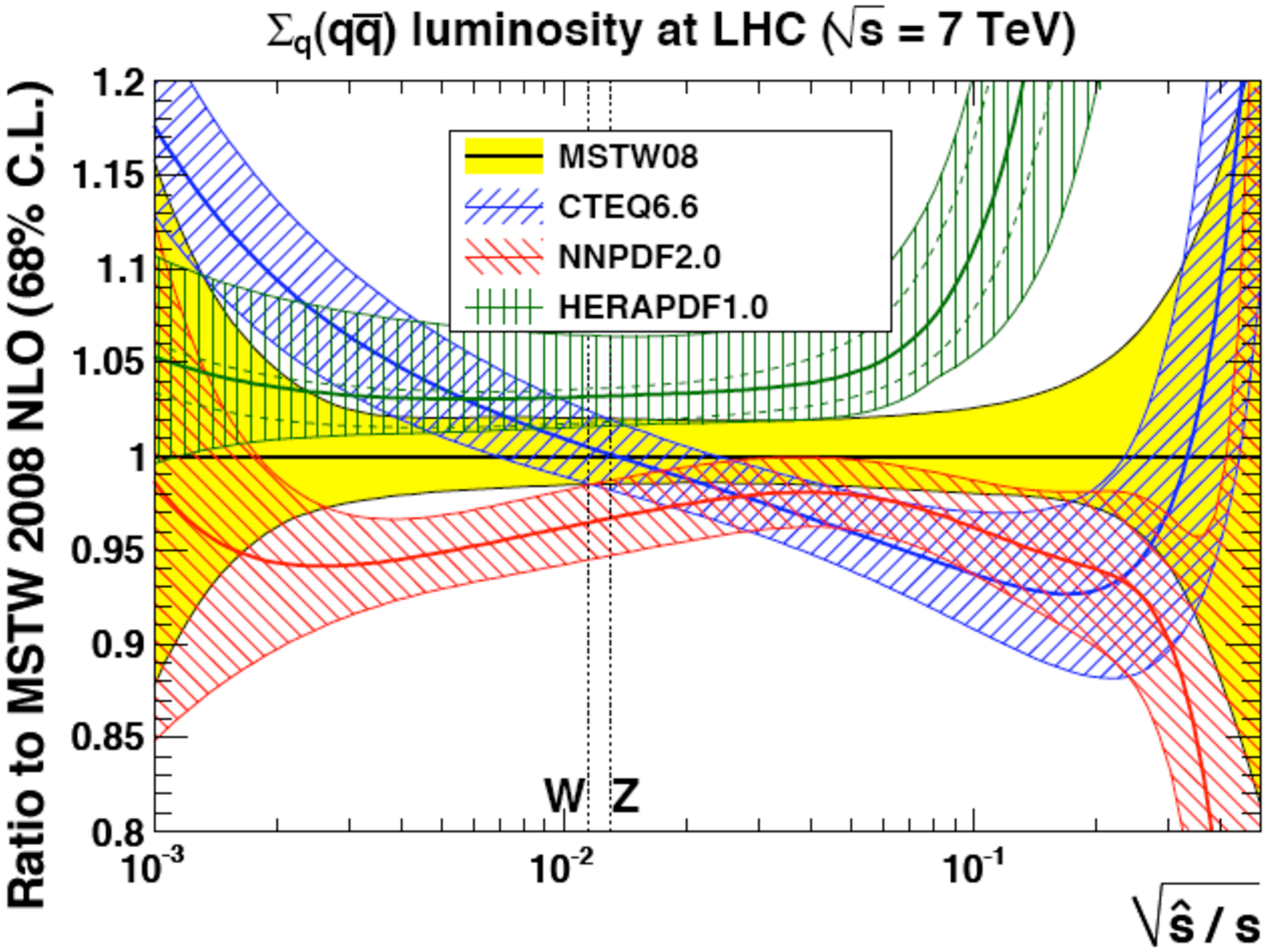}
\includegraphics[width=0.48\textwidth]{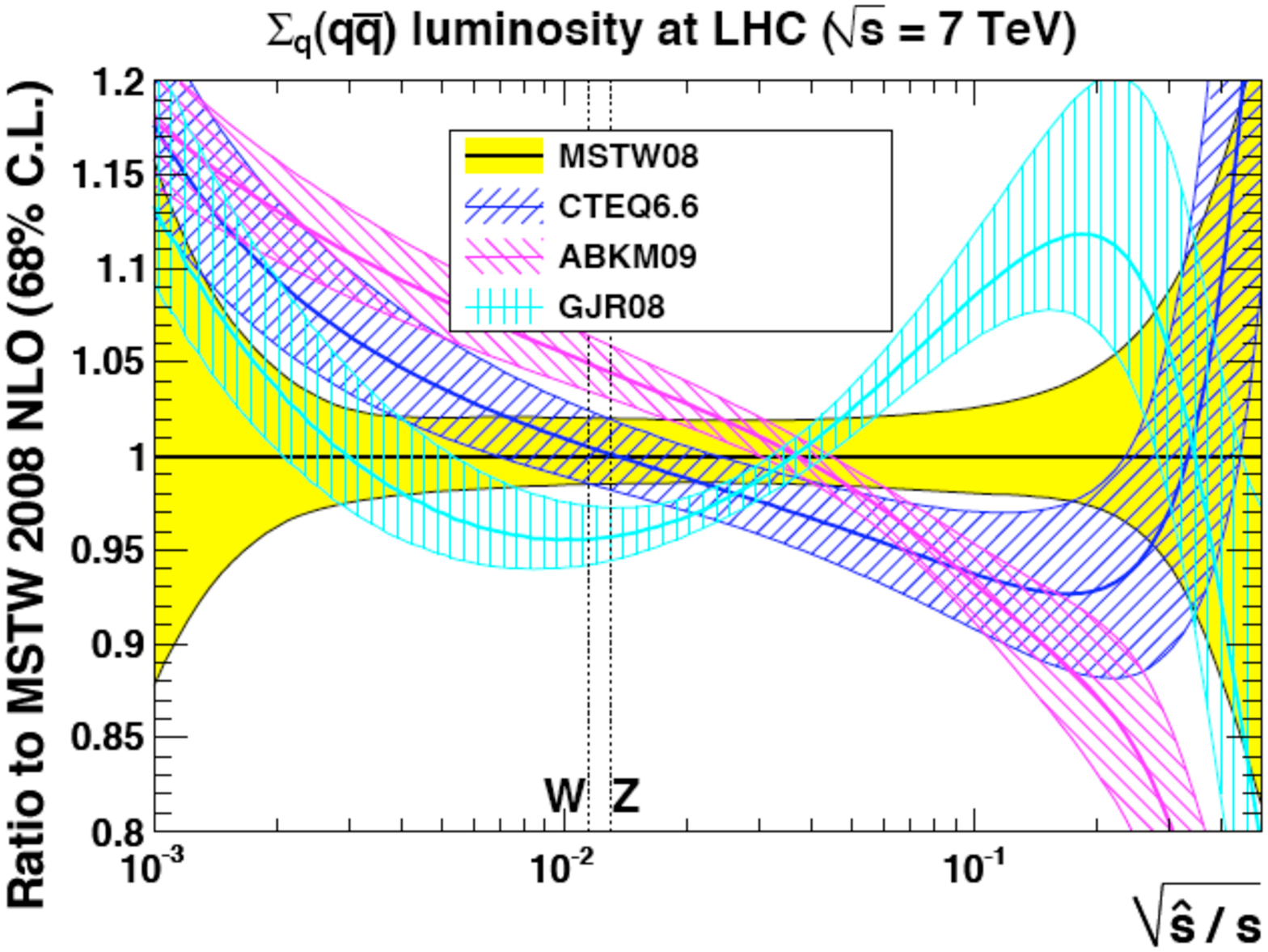}
\caption{\label{fig:qqlum} The $q\bar{q}$ luminosity functions and
  their uncertainties at 7 TeV, normalized to the MSTW08 result. 
Plot by G. Watt \cite{Watt}.}
\end{center}
\end{figure}

\begin{figure}
\begin{center}
\includegraphics[width=0.48\textwidth]{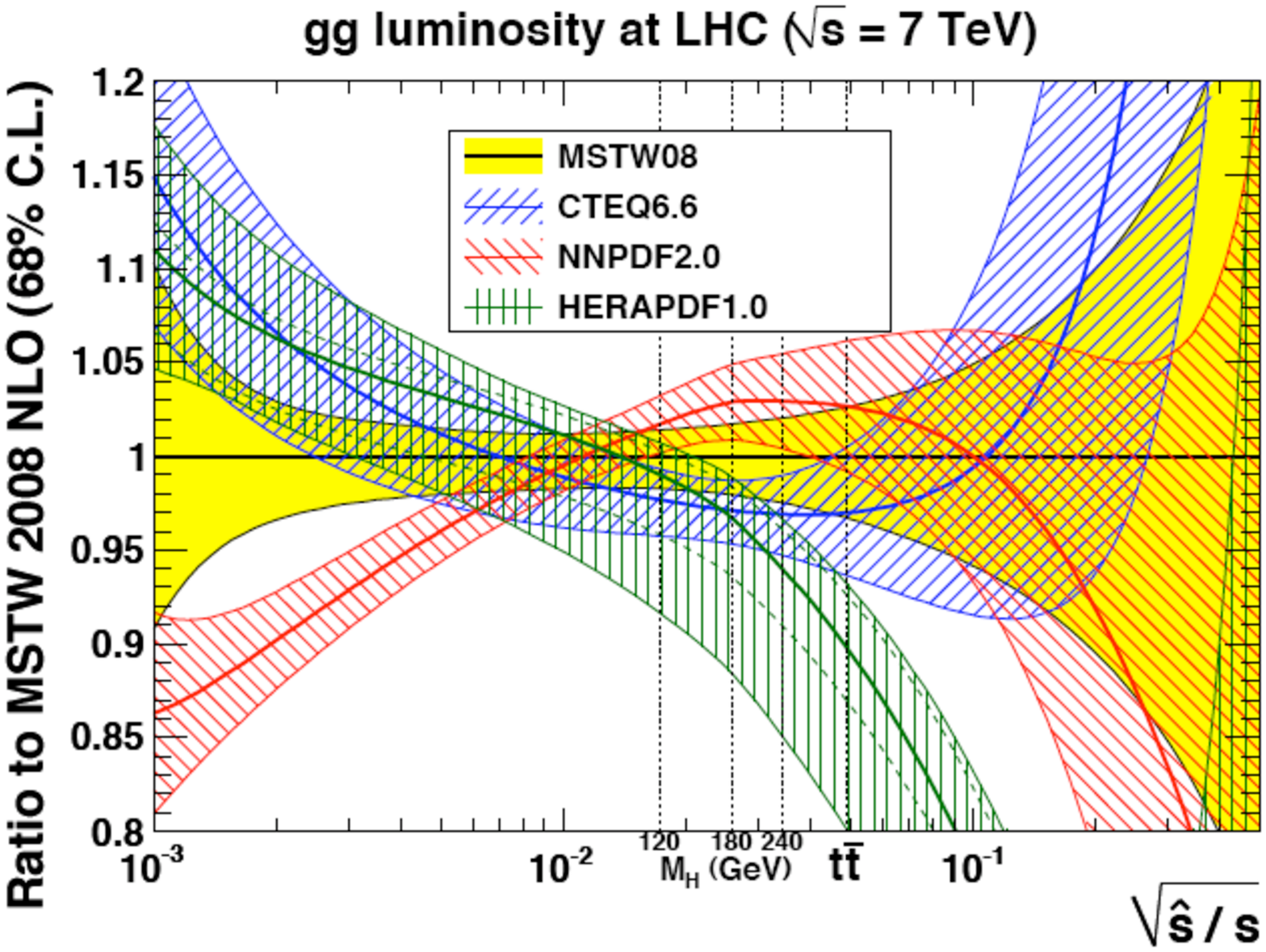}
\includegraphics[width=0.48\textwidth]{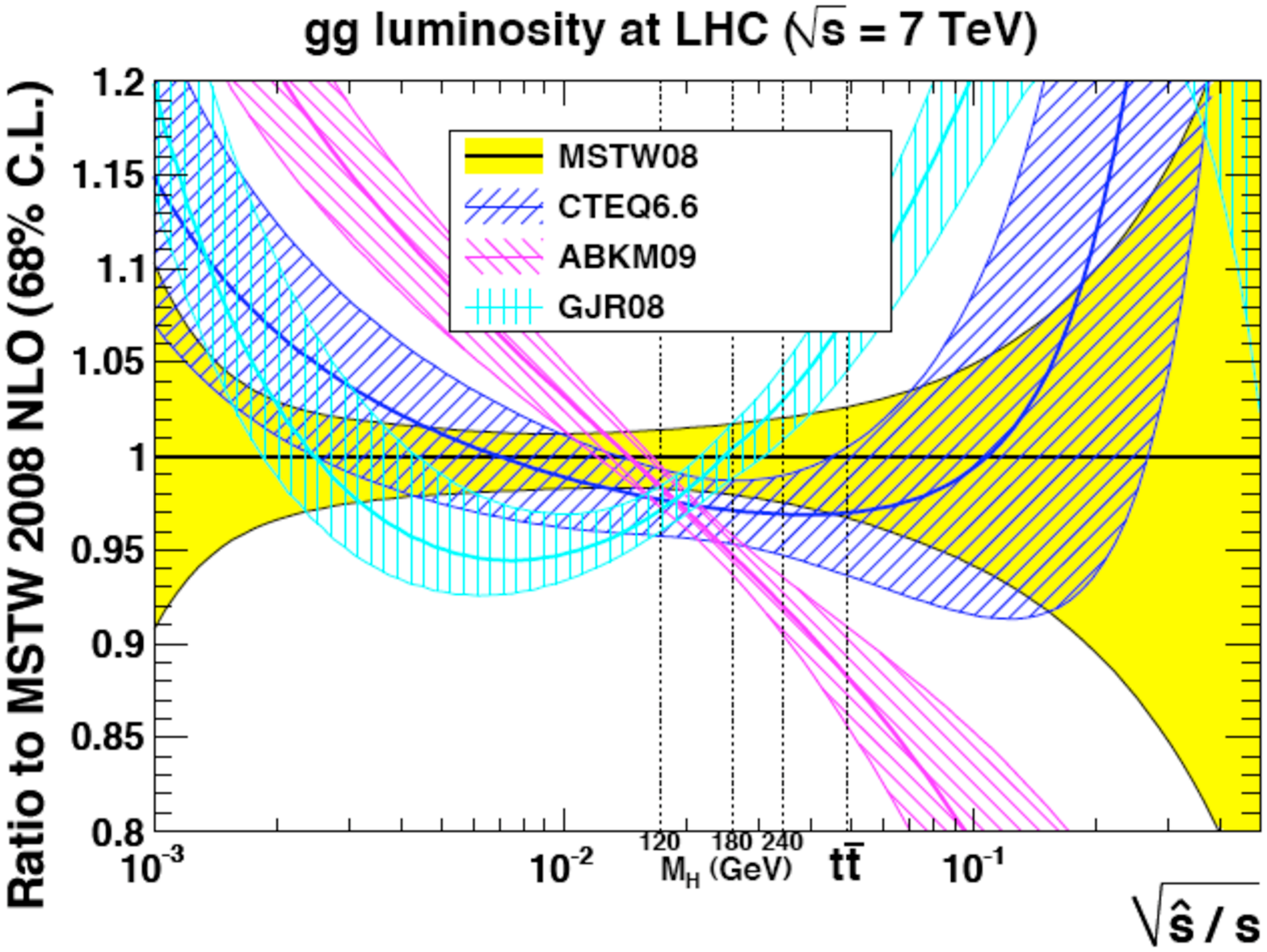}
\caption{\label{fig:gglum} The $gg$ luminosity functions and their
  uncertainties at 7 TeV, normalized to the MSTW08 result.
Plot by G. Watt \cite{Watt}.}
\end{center}
\end{figure}

Relative quark-antiquark and gluon-gluon PDF
luminosities are shown in Figures~\ref{fig:qqlum} and \ref{fig:gglum}.
CTEQ6.6,
NNPDF2.0, HERAPDF1.0, MSTW08, ABKM09 and GJR08 PDF luminosities are shown, all
normalized to the MSTW08 central value,
along with their 68 \%c.l.
error bands. The inner uncertainty bands (dashed lines)for HERAPDF1.0 
correspond to the (asymmetric) experimental errors, while the
outer uncertainty bands (shaded regions) also includes the model and
parameterisation errors. It is
interesting to note that the error bands for each of the PDF
luminosities are of  similar size. The
predictions of W/Z, $t\bar{t}$ and Higgs cross sections  are in
reasonable agreement for CTEQ, MSTW and NNPDF, while the agreement
with ABKM, HERAPDF and GJR is somewhat worse.
(Note however that these plots do not illustrate the effect that 
the different $\alpha_s(m_Z)$ values used by different groups will have 
on (mainly) $t\bar{t}$ and Higgs cross sections.)
It is also notable that the PDF
luminosities tend to differ at low $x$ and high $x$, for both
$q\bar{q}$ and $gg$ luminosities. The CTEQ6.6 distributions, for
example, may be larger at low $x$ than MSTW2008, due to the
positive-definite parameterization of the gluon distribution; the MSTW
gluon starts off negative at low $x$ and $Q^2$ and this results in an
impact for both the gluon and sea quark distributions at larger $Q^2$
values. The NNPDF2.0 $q\bar{q}$ luminosity tends to be somewhat lower,
in the $W,Z$ region for example.  Part of this effect might come from
the use of a ZM heavy quark scheme, although other differences might 
be relevant. 

\begin{figure}
\begin{center}
\includegraphics[width=0.48\textwidth]{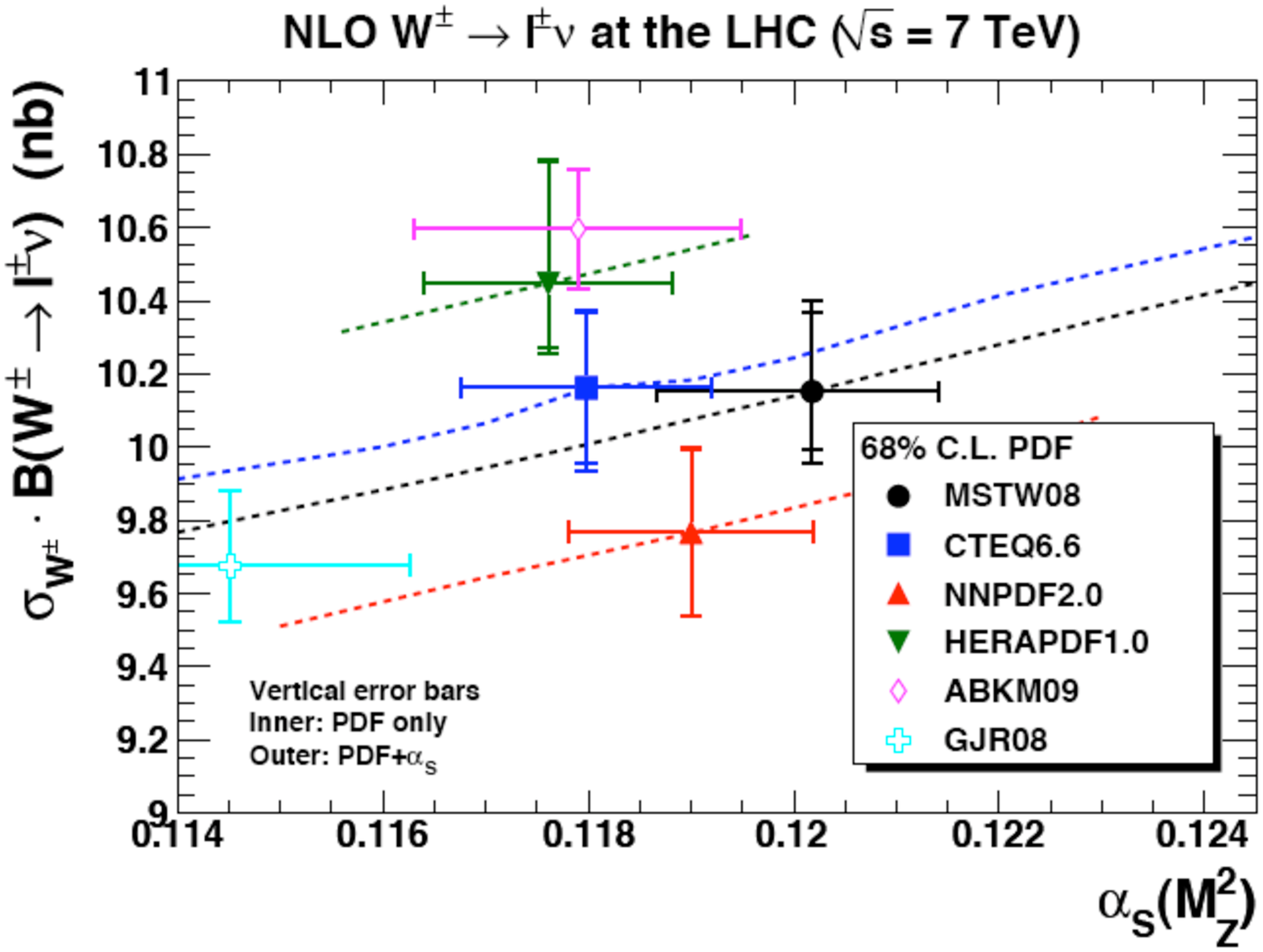}
\includegraphics[width=0.48\textwidth]{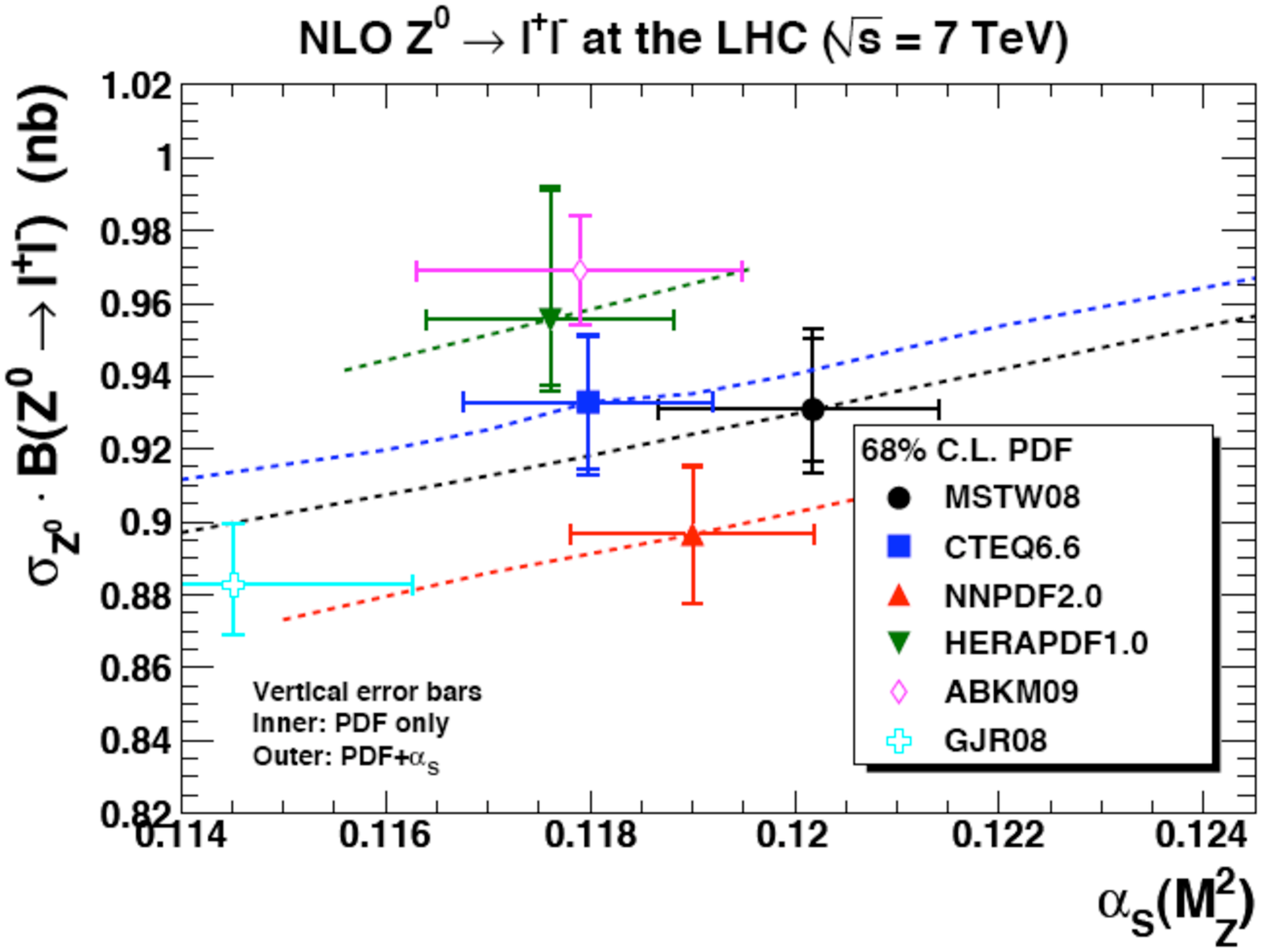}
\caption{\label{fig:WZ1} Cross section predictions at 7 TeV for $W^\pm$ and $Z$ production. All $Z$ cross sections plotted here
use a value of $\sin^2\theta_W=0.23149$. Plot by G. Watt \cite{Watt}.}
\end{center}
\end{figure}

\begin{figure}

\begin{center}
\includegraphics[width=0.48\textwidth]{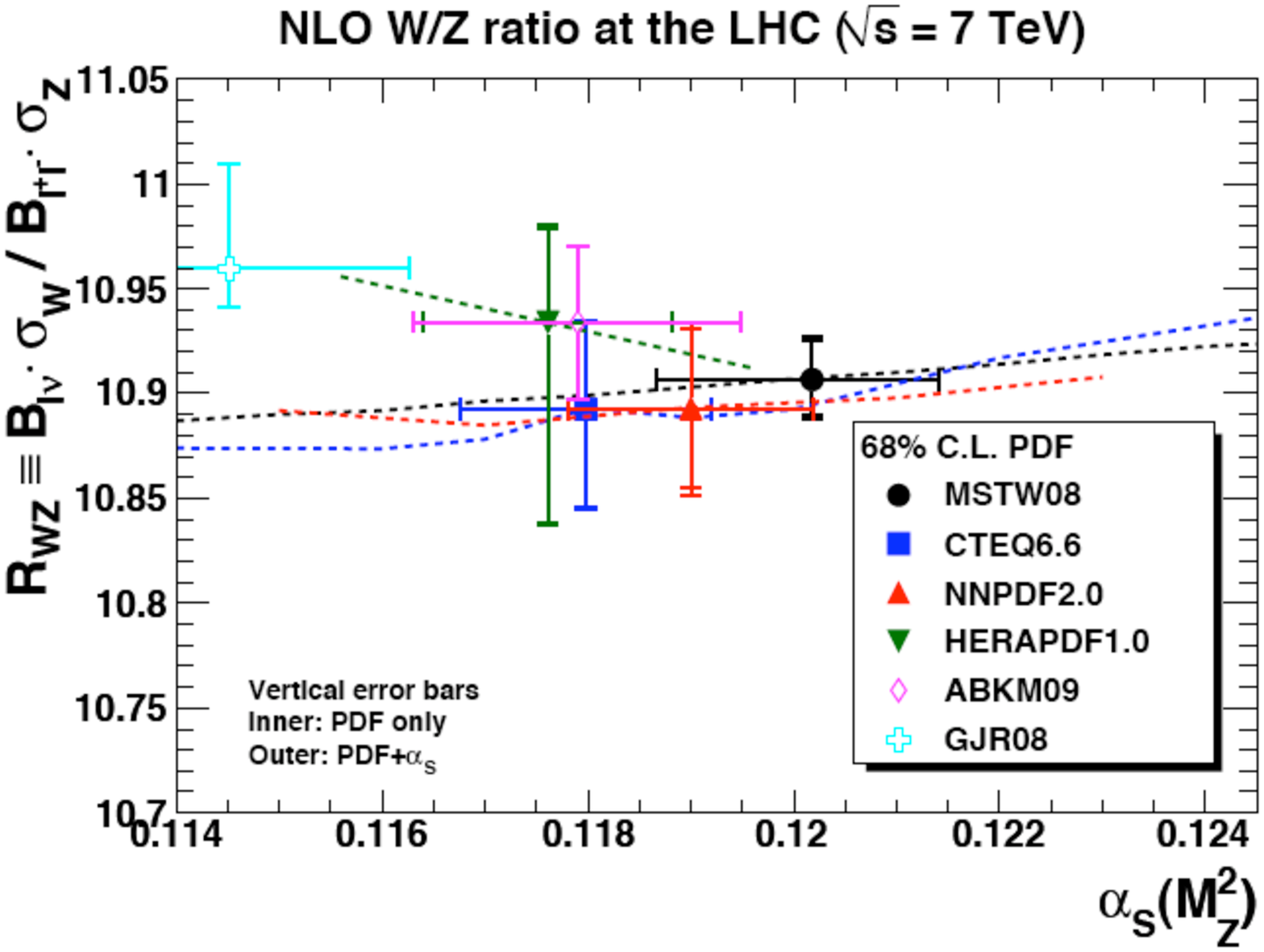}
\includegraphics[width=0.48\textwidth]{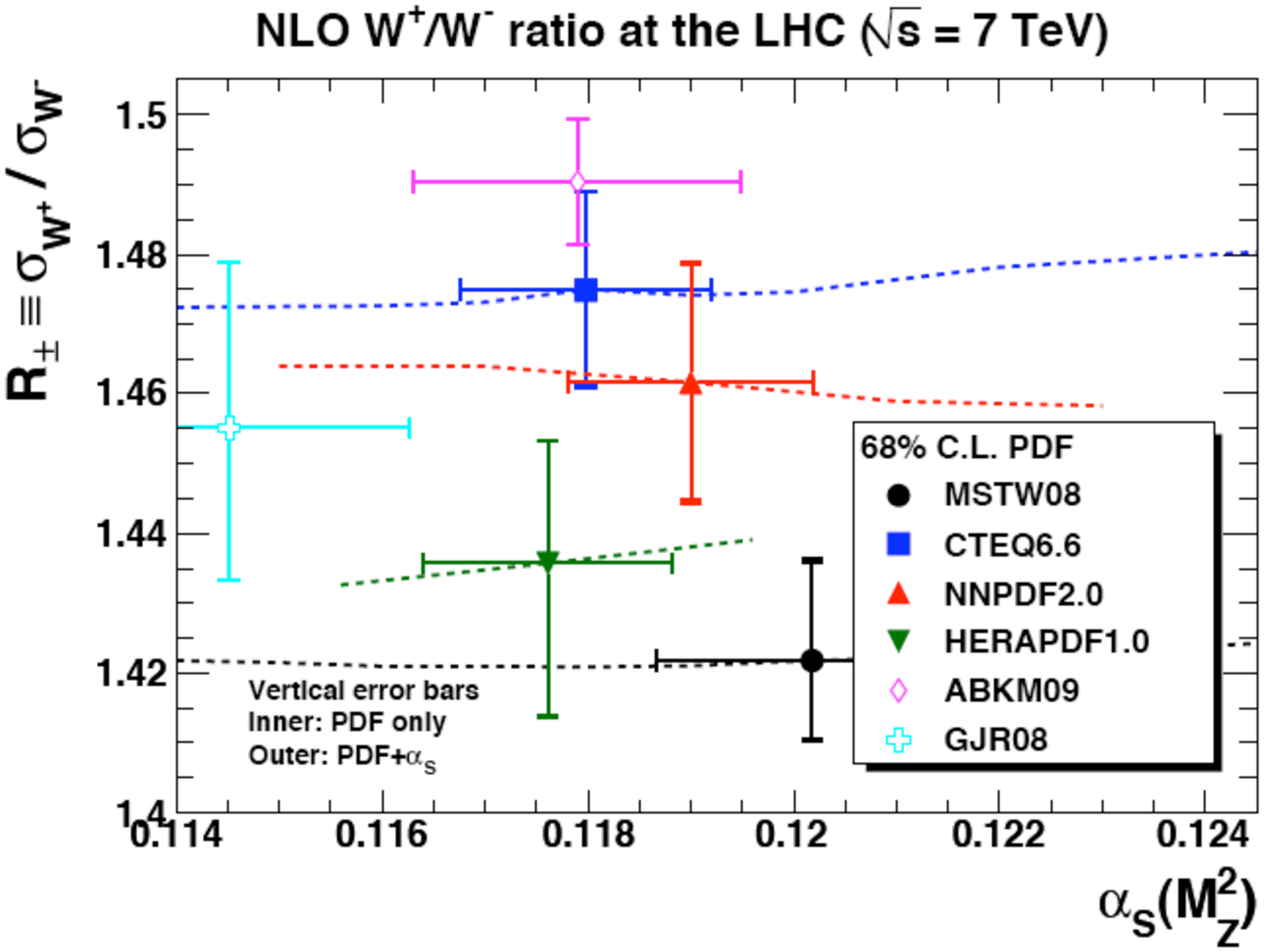}
\caption{\label{fig:WZ2} Cross section predictions at 7 TeV for the $W/Z$ and $W^{+}/W^{-}$ production. All $Z$ cross sections plotted here
use a value of $\sin^2\theta_W=0.23149$. Plot by G. Watt \cite{Watt}.}
\end{center}
\end{figure}

\begin{figure}
\begin{center}
\includegraphics[width=0.48\textwidth]{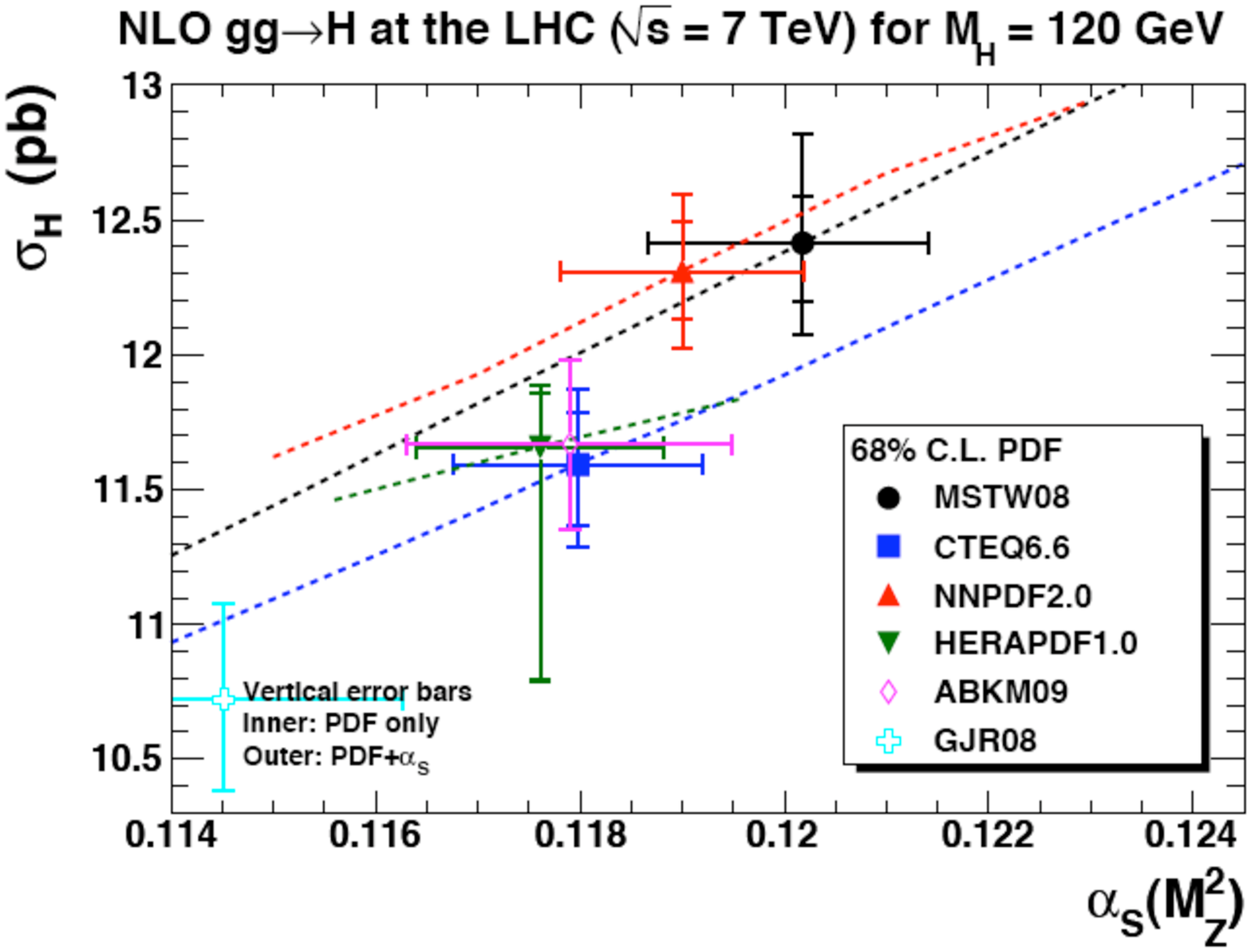}
\includegraphics[width=0.48\textwidth]{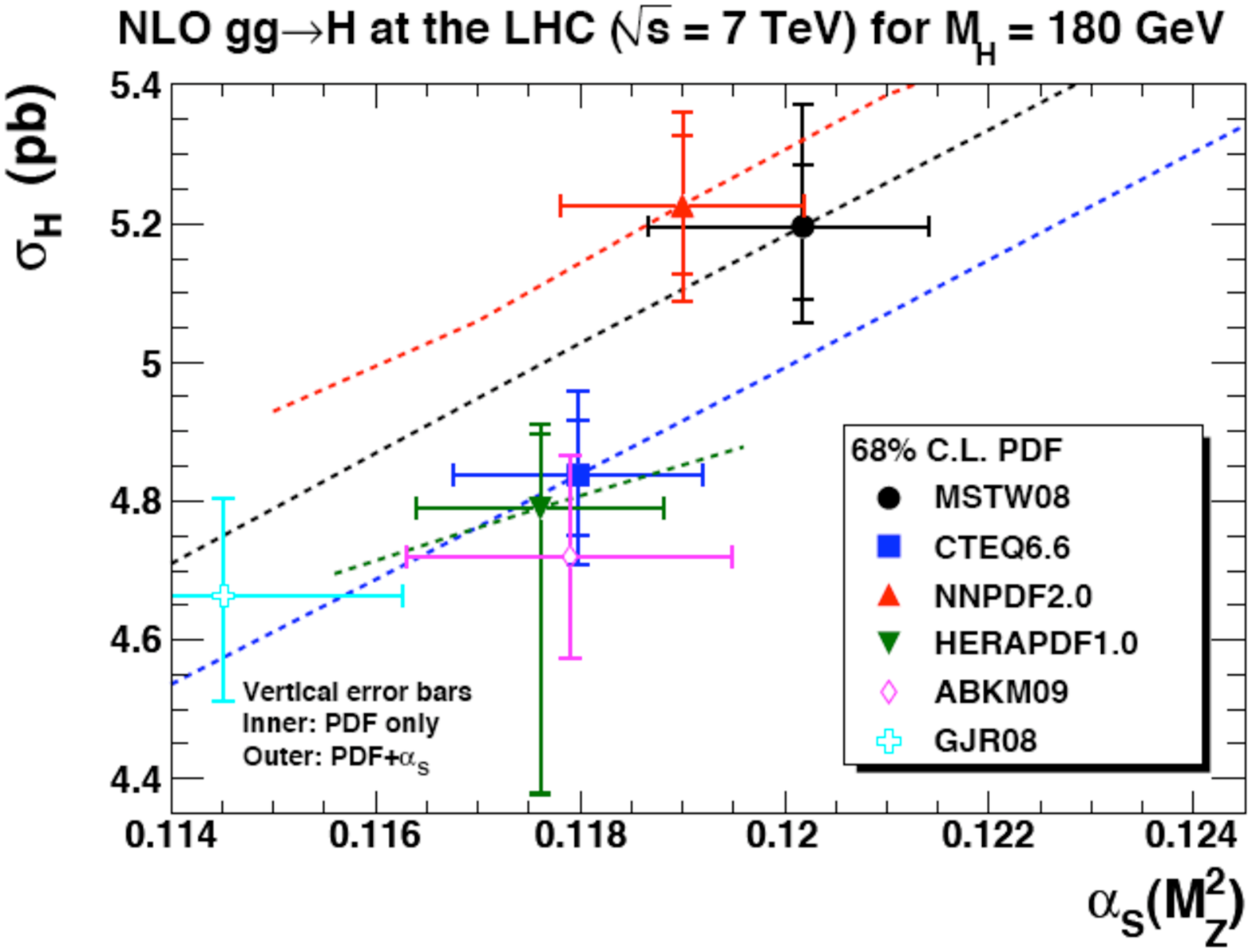}
\caption{\label{fig:higgs1} Cross section predictions at 7 TeV for a Higgs boson ($gg$ fusion) for a Higgs mass of 120 GeV (left) and 180 GeV(right).
Plot by G. Watt \cite{Watt}.}
\end{center}
\end{figure}

\begin{figure}
\begin{center}
\includegraphics[width=0.48\textwidth]{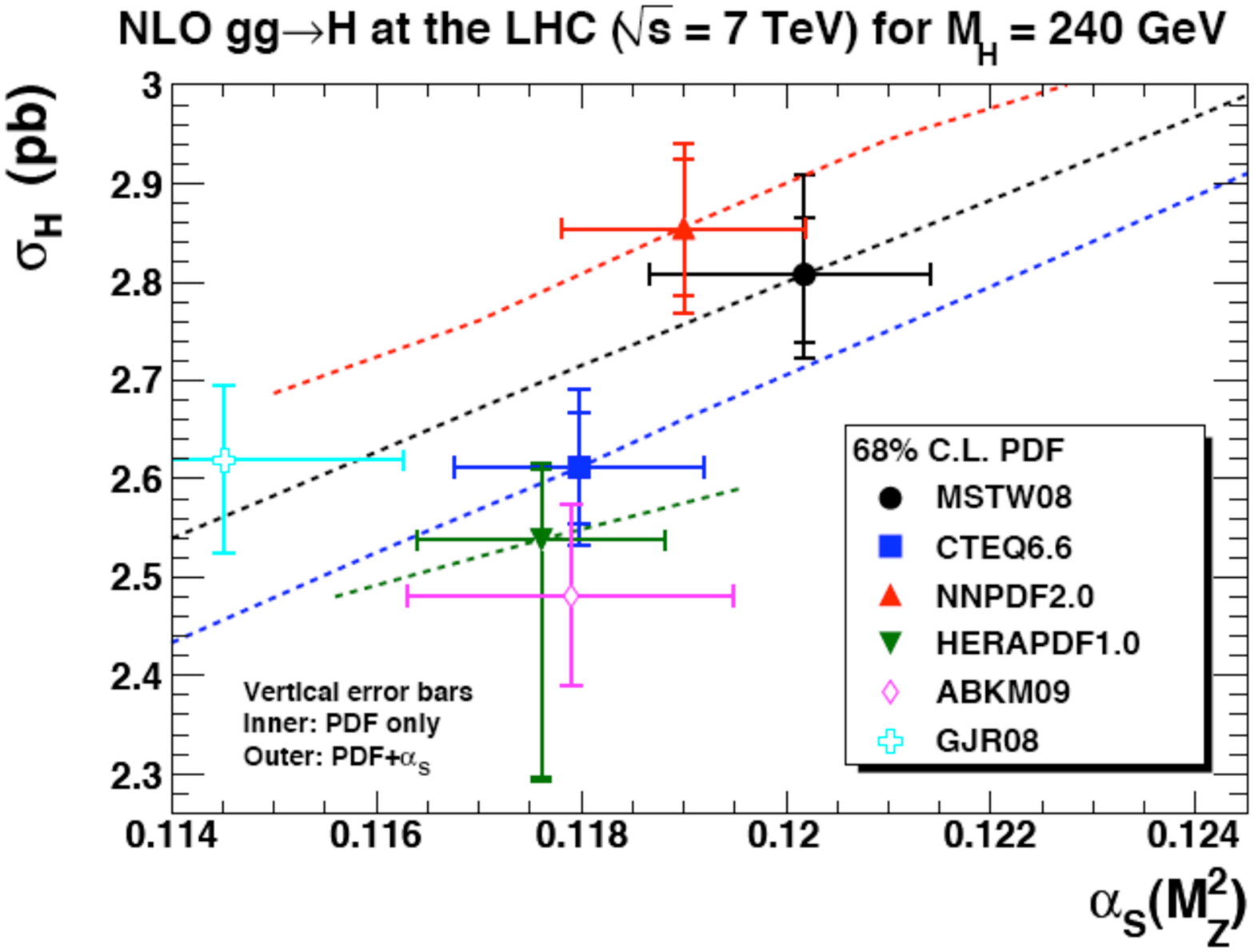}
\includegraphics[width=0.48\textwidth]{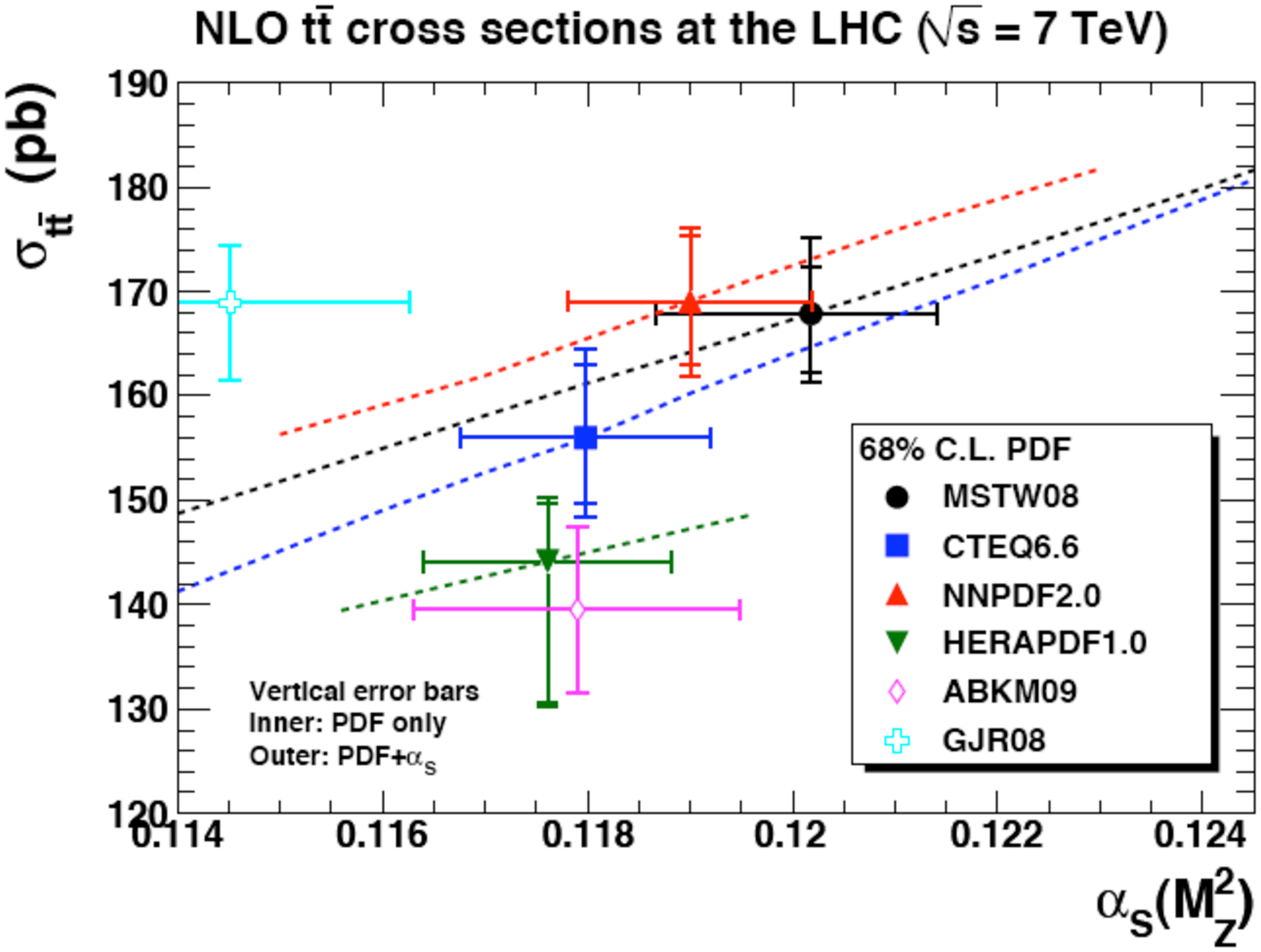}
\caption{\label{fig:higgs2} Cross section predictions at 7 TeV for a Higgs boson of mass 240 GeV (left) and for $t\bar{t}$ production (right).
Plot by G. Watt \cite{Watt}.}
\end{center}
\end{figure}

After having performed the comparison between PDF luminosities, we turn
to the comparison of LHC observables. Perhaps the most useful manner
to perform this comparison is to show the cross--sections as a
function of $\alpha_s$, with an interpolating curve connecting
different values of $\alpha_s$ for the same group, when
available~\cite{Watt} (see
Figs.~\ref{fig:WZ1}-\ref{fig:higgs2}). Following the interpolating
curve, it is possible to compare
cross sections at the same value of $\alpha_s$.
The predictions for the CTEQ, MSTW and NNPDF $W$ and $Z$ cross
sections at 7 TeV (Figs.~\ref{fig:WZ1}-\ref{fig:WZ2})
agree well, with  the NNPDF predictions somewhat lower, consistent
with the behaviour of the luminosity observed in Fig.~\ref{fig:qqlum}.
The cross sections from HERAPDF1.0 and ABKM09 are somewhat
larger~\footnote{Updated
versions of these plots, including an extension to NNLO, 
will be presented in a forthcoming MSTW publication.  See also Ref.~\cite{Alekhin:2010dd}.}.
The impact from the variation of the value of $\alpha_s$ is
relatively small. Basically, all of the PDFs
predict similar values for the $W/Z$ cross section ratio; much of the remaining uncertainty in this ratio is related to uncertainties in the strange quark distribution. This will
serve as a useful benchmark at the LHC. A larger variation in
predictions can be observed for the $W^+/W^-$ ratio (see
Fig.~\ref{fig:WZ2}).   
This quantity depends on the separation of the quarks into flavours 
and the separation between quarks and antiquarks. The data providing this 
information only extends down to $x=0.01$, and consists partially of 
neutrino DIS off nuclear targets. Hence, different groups provide different 
results because they fit different choices of data, make different 
assumptions about nuclear corrections and make 
different assumptions about the parametric forms of nonsinglet quarks 
relevant for $x \leq 0.01$. 

The predictions for Higgs production  from $gg$ fusion
(Figs.~\ref{fig:higgs1}-\ref{fig:higgs2}) depend strongly on the value
of $\alpha_s$: the anticorrelation between the gluon distribution and
the value of $\alpha_s$ is not sufficient to offset the growth of the
cross section (which starts at $O(\alpha_s^2)$ and undergoes a large
$O(\alpha_s^3)$  correction).
The CTEQ, MSTW and NNPDF predictions are in moderate agreement
but CTEQ lies somewhat lower, to some extent due to
the lower choice of $\alpha_s(M_Z^2)$. Compared at the common value of 
$\alpha_s(M_Z^2)=0.119$, the CTEQ prediction and that of either MSTW or
NNPDF, have one-sigma PDF uncertainties which just about overlap for each
value of $m_H$. If the comparison is made at the respective reference 
values of $\alpha_s$, but without accounting for the $\alpha_s$ uncertainty,
the discrepancies are rather worse, and indeed, even allowing for $\alpha_s$
uncertainty, the bands do not overlap. Hence, both the difference in PDFs and
in the dependence of the cross section on the value of $\alpha_s$ are
responsible for the differences observed. A useful measure of this 
is to note that the difference in the central values of the MSTW and 
CTEQ predictions for a common value of $\alpha_s(M_Z^2)=0.119$ for a 
120 GeV Higgs (a typical discrepancy) is equivalent to a change in 
$\alpha_s(M_Z^2)$ of about 0.0025. The worst PDF discrepancy is similar
to a change of about 0.004.
The predictions from HERAPDF  are rather lower, reflecting the
behaviour of the gluon luminosity of Fig.~\ref{fig:gglum}. The ABKM
and GJR predictions are also rather lower, but the $\alpha_s$
dependence of results is not explicitly available for these groups, hence it is
hard to tell how much of the discrepancy is due to the fact that these
groups adopt low values of $\alpha_s$.

Production of a $t\bar{t}$ pair (Fig.~\ref{fig:higgs2}, right plot)
probes the gluon-gluon luminosity at a
higher value of $\sqrt{\hat{s}}$, with smaller higher order
corrections than present for Higgs production through $gg$ fusion. The
cross section predictions from CTEQ6.6, MSTW2008 and NNPDF2.0 are all
seen to be in good agreement, especially when 
evaluated at the common value of $\alpha_s(m_Z)$ of
0.119.  

\subsection{Tables of results from each PDF set}

In the subsections below, we provide tables of the benchmark cross
sections from the PDF groups participating in the benchmark
exercise. Only results for 7 TeV will be provided for this interim version of the note.

\subsubsection{ABMK09 NLO 5 Flavours}

In the following sub-section, the tables of relevant cross sections for the ABKM09 PDFs are given. Results are given for the value of $\alpha_s(m_Z)$ determined from the fit. The charm mass is taken to be $1.5 \pm 0.25$ GeV and the bottom mass is taken to be  $4.5 \pm 0.5$ GeV. The heavy quark mass uncertainites are incorporated in with the PDF uncertainties.

The results obtained with the ABKM09 NLO 5 flavours 
set are reported in
Tables~1-2.


  \begin{center}    
    \begin{tabular}{|l||c|c|c|c|c|c|}
      \hline
{\small Process} & {\small Cross section} & {\small combined PDF and $\alpha_s$ errors } \\
	\hline
{\small $\sigma_{W^+}*BR(W^+\rightarrow l^+\nu)[nb]$}  & 6.3398 &    0.0981 \\
{\small $\sigma_{W^-}*BR(W^-\rightarrow l^-\nu)[nb]$ } &  4.2540 &   0.0657\\
{\small $\sigma_{Z^o}*BR(Z^o\rightarrow l^+ l^-)[nb]$}  &  0.9834 &  0.0151 \\
{\small $\sigma_{t\bar{t}}[pb]$} & 139.55  & 7.96 \\
{\small $\sigma_{gg\rightarrow Higgs}(120~ GeV)[pb]$ } &  11.663 &   0.314\\
{\small $\sigma_{gg\rightarrow Higgs}(180~ GeV)[pb]$ } &  4.718 &    0.147\\
{\small $\sigma_{gg\rightarrow Higgs}(240~ GeV)[pb]$}  &  2.481 &    0.092\\

\hline
\end{tabular}
\end{center}
Table 1. Benchmark cross section predictions and uncertainties
  for 
  ABKM09 NLO $n_f=5$
  for $W^{\pm},Z, t\bar{t}$ and Higgs production
  (120, 180, 240 GeV) at 7 TeV.
  The central prediction is given in column 2. 
Errors are quoted at the 68\% CL. The PDF
 and $\alpha_s(m_Z)$ errors are evaluated simultaneously. Higgs boson cross sections are corrected for finite top mass effects (1.06, 1.15 and 1.31 for masses of 120, 180 and 240 GeV respectively.

  \begin{center}    
    \begin{tabular}{|l||c|c|c|c|c|c|c|}
      \hline
      $ y_W $ & ${d\sigma_{W^+}\over{dy}}*BR$ & PDF + $\alpha_s$ Error & ${d\sigma_{W^-}\over{dy}}*BR$ & PDF + $\alpha_s$ Error &${d\sigma_{Z^o}\over{dy}}*BR$ & PDF + $\alpha_s$ Error \\
\hline
-4.4 & 0.002   & 0.0005  & 0.0001  & 0.00004  & 0.00002   &  0.000004  \\
-4.0 & 0.102   & 0.0084  & 0.0198  & 0.00262  & 0.00472   &  0.000324  \\
-3.6 & 0.394   & 0.0114  & 0.1228  & 0.01140  & 0.03321   &  0.000909  \\
-3.2 & 0.687   & 0.0324  & 0.2663  & 0.03815  & 0.07542   &  0.002259 \\
-2.8 & 0.878   & 0.0368  & 0.4017  & 0.04089  & 0.10946   &  0.002440 \\
-2.4 & 0.940   & 0.0298  & 0.5328  & 0.01768  & 0.13367   &  0.002566 \\
-2.0 & 0.935   & 0.0180  & 0.6249  & 0.01945  & 0.14787   &  0.002834 \\
-1.6 & 0.915   & 0.0215  & 0.6923  & 0.01479  & 0.15581   &  0.002905 \\
-1.2 & 0.895   & 0.0219  & 0.7344  & 0.01717  & 0.16042   &  0.004083  \\
-0.8 & 0.881   & 0.0241  & 0.7625  & 0.02627  & 0.16298   &  0.003530  \\
-0.4 & 0.867   & 0.0241  & 0.7729  & 0.02364  & 0.16373   &  0.004749  \\
0.0 &  0.863   & 0.0402  & 0.7774  & 0.02215  & 0.16463   &  0.003186   \\
0.4 &  0.870   & 0.0411  & 0.7733  & 0.01379  & 0.16352   &  0.005058   \\
0.8 &  0.871   & 0.0254  & 0.7603  & 0.01647  & 0.16260   &  0.003751   \\
1.2 &  0.891   & 0.0461  & 0.7348  & 0.02070  & 0.16092   &  0.003715  \\
1.6 &  0.926   & 0.0589  & 0.6920  & 0.01416  & 0.15539   &  0.004267   \\
2.0 &  0.934   & 0.0234  & 0.6255  & 0.01680  & 0.14750   &  0.003665   \\
2.4 &  0.938   & 0.0161  & 0.5279  & 0.01737  & 0.13373   &  0.003013  \\
2.8 &  0.873   & 0.0244  & 0.4045  & 0.01109  & 0.10944   &  0.002216  \\
3.2 &  0.692   & 0.0173  & 0.2658  & 0.00600  & 0.07541   &  0.001574   \\
3.6 &  0.393   & 0.0123  & 0.1254  & 0.00765  & 0.03353   &  0.001316   \\
4.0 &  0.100   & 0.0057  & 0.0178  & 0.00434  & 0.00441   &  0.000361   \\
4.4 &  0.002   & 0.0004  & 0.0001  & 0.00003  & 0.00001   &  0.000003   \\
\hline
\end{tabular}
\end{center}
Table 2. Benchmark cross section predictions ($d\sigma/dy*BR$ in nb)
 for ABKM09 NLO with $n_f=5$ for $W^{\pm},Z^o$ production at 
7 TeV, as a function of boson rapidity.


\subsubsection{CTEQ6.6}
\label{sec:cteq66}

In the following sub-section, the tables of relevant cross sections for the CTEQ6.6 PDFs are given (Tables 3-6). The predictions for the central value of $\alpha_s(m_Z)$ are given in bold. Errors are quoted at the 68\% c.l. For CTEQ6.6, this involves dividing the normal 90\%c.l. errors by a factor of 1.645.

  \begin{center}    
    \begin{tabular}{|l||c|c|c|}
      \hline
      {\small $\alpha_s(m_Z)$ }& {\small $\sigma_{W^+}*BR(W^+\rightarrow l^+\nu$)[nb]} &{\small $\sigma_{W^-}*BR(W^-\rightarrow l^-\nu)[nb]$ } & {\small $\sigma_{Z^o}*BR(Z^o\rightarrow l^+ l^-)[nb]$}   \\
	\hline
0.116 &  5.957 & 4.044 & 0.9331  \\
0.117 &  5.993 & 4.068 & 0.9384  \\
$\bf 0.118$ &  $\bf 6.057$ & $\bf 4.106 $ & $\bf 0.9469 $  \\

0.119 &  6.064 & 4.114 & 0.9485  \\
0.120 &  6.105 & 4.139 & 0.9539  \\
      \hline
    \end{tabular}
  \end{center}
Table 3: Benchmark cross section predictions for CTEQ6.6 for $W^{\pm},Z$ and $t\bar{t}$ production at 7 TeV, as a function of $\alpha_s(m_Z)$. The results for the central value of $\alpha_s(m_Z)$ for CTEQ6.6 (0.118) are shown in bold.  

  \begin{center}    
    \begin{tabular}{|l||c|c|c|c|}
      \hline
      $\alpha_s(m_Z)$ & $\sigma_{gg\rightarrow Higgs}(120~ GeV)[pb]$ & $\sigma_{gg\rightarrow Higgs}(180~ GeV)[pb]$& $\sigma_{gg\rightarrow Higgs}(240~ GeV)[pb]$ & {\small $\sigma_{t\bar{t}}[pb]$}\\
	\hline
0.116 & 11.25  & 4.69  & 2.52  & 149.2\\
0.117 & 11.42  & 4.76  & 2.57 & 153.0\\
$\bf 0.118$ &  $\bf 11.59$ & $\bf 4.84 $ & $\bf 2.61 $ & $ \bf 156.2 $  \\
0.119 & 11.75  & 4.91 & 2.66 & 160.5 \\
0.120 & 11.92  & 4.99 & 2.70 & 164.3  \\
      \hline
    \end{tabular}
  \end{center}
Table 4: Benchmark cross section predictions for CTEQ6.6 for $gg\rightarrow Higgs$ production (masses of 120, 180 and 240 GeV), and for $t\bar{t}$ production,  at 7 TeV, as a function of $\alpha_s(m_Z)$. The results for the central value of $\alpha_s(m_Z)$ for CTEQ6.6 (0.118) are shown in bold.  Higgs production ross sections have been corrected for the finite top mass effect (a factor of 1.06 for 120 GeV, 1.15 for 180 GeV and 1.31 for 240 GeV). 

  \begin{center}    
    \begin{tabular}{|l||c|c|c|c|c|c|c|}
      \hline
{\small Process} & {\small $\sigma$} & {\small PDF (asym)} & {\small PDF (sym)} & {\small $\alpha_s(m_Z)$ error} & {\small combined} & \small correlation \\
	\hline
{\footnotesize $\sigma_{W^+}*BR(W^+\rightarrow l^+\nu)[nb]$}  &  6.057 & +0.123/-0.119 & 0.116 & 0.045 & 0.132 & 0.87 \\
{\footnotesize $\sigma_{W^-}*BR(W^-\rightarrow l^-\nu)[nb]$ } &  4.106 & +0.088/-0.091 & 0.088 & 0.029 & 0.092 & 0.92\\
{\footnotesize $\sigma_{Z^o}*BR(Z^o\rightarrow l^+ l^-)[nb]$}  &  0.9469  &  +0.018/-0.018 &  0.018 & 0.006 & 0.0187 & 1.00 \\
$\sigma_{t\bar{t}}[pb]$ &  156.2 & +7.0/-6.7 & 6.63  & 4.59 & 8.06 & -0.74  \\
{\footnotesize $\sigma_{gg\rightarrow Higgs}(120~ GeV)[pb]$ } &  11.59 & +0.19/-0.23 & 0.21 & 0.20 & 0.29 & 0.01 \\
{\footnotesize $\sigma_{gg\rightarrow Higgs}(180~ GeV)[pb]$ } &  4.840 & +0.077/-0.091 & 0.084 & 0.091 & 0.124 & -0.47\\
{\footnotesize $\sigma_{gg\rightarrow Higgs}(240~ GeV)[pb]$}  &  2.610 & +0.054/-0.058 & 0.056 & 0.055 & 0.078 & -0.73 \\

\hline
\end{tabular}
\end{center}
Table 5: Benchmark cross section predictions and uncertainties for CTEQ6.6 for $W^{\pm},Z, t\bar{t}$ and Higgs production (120, 180, 240 GeV) at 7 TeV. The central prediction is given in column 2. Errors are quoted at the 68\% c.l.. Both the symmetric and asymmetric forms for the PDF errors are given. In the next-to-last column, the (symmetric) form of the PDF and $\alpha_s(m_Z)$ errors are added in quadrature.  In the last column, the correlation cosine with respect to $Z$ production is given. 

  \begin{center}    
    \begin{tabular}{|l||c|c|c|c|c|c|c|}
      \hline
      $ y_W $ & ${d\sigma_{W^+}\over{dy}}*BR$ & PDF Error & ${d\sigma_{W^-}\over{dy}}*BR$ & PDF Error &${d\sigma_{Z^o}\over{dy}}*BR$ & PDF Error \\
\hline
-4.4 &  0.002 & 0.0005 & 0.000 & 0.0000 & 0.000 & 0.0000 \\
-4.0 &  0.094 & 0.006 & 0.019 & 0.0063 & 0.005 & 0.00032 \\
-3.6 &  0.367 & 0.013 & 0.122 & 0.0126 & 0.031 & 0.00109  \\
-3.2 &  0.634 & 0.016 & 0.274 & 0.013 & 0.071 & 0.00184 \\
-2.8 &  0.806 & 0.0187 & 0.414 & 0.0128 & 0.106 & 0.00235 \\
-2.4 &  0.878 & 0.019 & 0.517 & 0.0131 & 0.127 & 0.00255 \\
-2.0 &  0.886 & 0.018 & 0.597 & 0.0134 & 0.141 & 0.00255 \\
-1.6 &  0.883 & 0.018 & 0.653 & 0.0144 & 0.148 &  0.00286 \\
-1.2 &  0.867 & 0.020 & 0.697 & 0.017 & 0.155 & 0.00347 \\
-0.8 &  0.862 & 0.023 & 0.723 & 0.02 & 0.166 & 0.00408 \\
-0.4 &  0.855 & 0.025 & 0.739 & 0.023 & 0.161 & 0.00469 \\
0.0 &  0.864  & 0.026 & 0.750 & 0.0236 & 0.162 & 0.0049  \\
0.4 &  0.854 & 0.025 & 0.740 & 0.0226 & 0.161 & 0.00479  \\
0.8 &  0.865 & 0.023 & 0.728 & 0.020& 0.158 & 0.00418  \\
1.2 &  0.870 & 0.020 & 0.690 & 0.0167 & 0.155 & 0.00347 \\
1.6 &  0.882 & 0.018 & 0.654 & 0.0144 & 0.148 & 0.00286  \\
2.0 &  0.890 & 0.018 & 0.606 & 0.0134 & 0.141 & 0.00265  \\
2.4 &  0.872 & 0.019 & 0.508 & 0.0128 & 0.114 & 0.0025 \\
2.8 &  0.806 & 0.019 & 0.416 & 0.0128 & 0.106 & 0.00235 \\
3.2 &  0.640 & 0.016 & 0.274 & 0.0128 & 0.071 & 0.00184  \\
3.6 &  0.364 & 0.013 & 0.120 & 0.0127 & 0.031 & 0.00109  \\
4.0 &  0.095 & 0.006 & 0.023 & 0.0064 & 0.005 & 0.00031  \\
4.4 &  0.003 & 0.0005 & 0.000 & 0.000 & 0.000 & 0.0000  \\

\hline
\end{tabular}
\end{center}
Table 6: Benchmark cross section predictions ($d\sigma/dy*BR$ in nb) for CTEQ6.6 for $W^{\pm},Z^o$ production at 7 TeV, as a function of boson rapidity. 



\subsubsection{GJR}

In the following sub-section, the tables of relevant cross sections for the GJR08 PDFs are given (Tables 7-8). The results are given at the fit value of $\alpha_s(m_Z)$ and the errors correspond to the PDF-only errors at 68\% c.l. 

  \begin{center}    
    \begin{tabular}{|l||c|c|c|}
      \hline
{\small Process} & {\small Cross section} & {\small PDF Error}  \\
	\hline
{\small $\sigma_{W^+}*BR(W^+\rightarrow l^+\nu)[nb]$}  &  5.74 & 0.11 \\
{\small $\sigma_{W^-}*BR(W^-\rightarrow l^-\nu)[nb]$ } &  3.94 & 0.08 \\
{\small $\sigma_{Z^o}*BR(Z^o\rightarrow l^+ l^-)[nb]$}  &  0.897  & 0.014 \\
$\sigma_{t\bar{t}}[pb]$ &  169 & 6 \\
{\small $\sigma_{gg\rightarrow Higgs}(120~ GeV)[pb]$ } &  10.72 & 0.35 \\
{\small $\sigma_{gg\rightarrow Higgs}(180~ GeV)[pb]$ } &  4.66 & 0.14 \\
{\small $\sigma_{gg\rightarrow Higgs}(240~ GeV)[pb]$}  &  2.62 & 0.09 \\

\hline
\end{tabular}
\end{center}

Table 7: Benchmark cross section predictions and uncertainties for GJR08 for $W^{\pm},Z, t\bar{t}$ and Higgs production (120, 180, 240 GeV) at 7 TeV. The results are given at the fit value of $\alpha_s(m_Z)$ and the errors correspond to the PDF-only errors at 68\% c.l. Higgs boson cross sections have been corrected for the finite top mass effect (a factor of 1.06 for 120 GeV, 1.15 for 180 GeV and 1.31 for 240 GeV). 

  \begin{center}    
    \begin{tabular}{|l||c|c|c|c|c|c|c|}
      \hline
      $ y_W $ & ${d\sigma_{W^+}\over{dy}}*BR$ & PDF Error & ${d\sigma_{W^-}\over{dy}}*BR$ & PDF Error &${d\sigma_{Z^o}\over{dy}}*BR$ & PDF Error \\
\hline
-4.4 &  0.002 & 0.000 & 0.000 & 0.000 & 0.000 & 0.000 \\
-4.0 &  0.091 & 0.003 & 0.018 & 0.004 & 0.004 & 0.000 \\

-3.6 &  0.368 & 0.011 & 0.122 & 0.013 & 0.031 & 0.001  \\
-3.2 &  0.640 & 0.016 & 0.275 & 0.016 & 0.071 & 0.002 \\
-2.8 &  0.789 & 0.018 & 0.424 & 0.016 & 0.106 & 0.002 \\
-2.4 &  0.848 & 0.018 & 0.525 & 0.016 & 0.126 & 0.002 \\
-2.0 &  0.844 & 0.017 & 0.596 & 0.015 & 0.137 & 0.002 \\
-1.6 &  0.831 & 0.016 & 0.633 & 0.014 & 0.141 &  0.002 \\
-1.2 &  0.803 & 0.015 & 0.654 & 0.013 & 0.144 & 0.002 \\
-0.8 &  0.785 & 0.015 & 0.668 & 0.013 & 0.143 & 0.002 \\
-0.4 &  0.777 & 0.014 & 0.672 & 0.013 & 0.144 & 0.003 \\
0.0 &  0.780  & 0.014 & 0.677 & 0.013 & 0.145 & 0.003  \\
0.4 &  0.777 & 0.014 & 0.673 & 0.013 & 0.144 & 0.003  \\
0.8 &  0.789 & 0.015 & 0.670 & 0.013 & 0.145 & 0.003  \\
1.2 &  0.806 & 0.015 & 0.655 & 0.013 & 0.143 & 0.002 \\
1.6 &  0.823 & 0.016 & 0.631 & 0.014 & 0.142 & 0.002  \\
2.0 &  0.852 & 0.017 & 0.596 & 0.015 & 0.137 & 0.002  \\
2.4 &  0.842 & 0.018 & 0.527 & 0.016 & 0.126 & 0.002 \\
2.8 &  0.791 & 0.018 & 0.422 & 0.016 & 0.106 & 0.002 \\
3.2 &  0.636 & 0.016 & 0.278 & 0.016 & 0.072 & 0.002  \\
3.6 &  0.371 & 0.011 & 0.117 & 0.013 & 0.031 & 0.001  \\
4.0 &  0.092 & 0.003 & 0.019 & 0.004 & 0.004 & 0.000  \\
4.4 &  0.023 & 0.000 & 0.000 & 0.000 & 0.000 & 0.000  \\

\hline
\end{tabular}
\end{center}
Table 8: Benchmark cross section predictions ($d\sigma/dy*BR$ in nb) for GJR for $W^{\pm},Z^o$ production at 7 TeV, as a function of boson rapidity. The results are given at the fit value of $\alpha_s(m_Z)$ and the errors correspond to the PDF errors at 68\% CL. 



\subsubsection{HERAPDF1.0}

In the following sub-section, the tables of relevant cross sections for the HERAPDF1.0 PDFs are given (Tables 9-10). The predictions for the central value of $\alpha_s(m_Z)$ are given in bold.



\begin{scriptsize}
\begin{center}
\begin{tabular}{|l|cccccc|}
\hline
Process & $\sigma$ & Exp. error & Model error  & Param. error & $\alpha_S$ error & Total error \\
\hline
$\sigma_{W^+}$ $\times BR(W^+ \to \ell^+\nu)$ [nb] & $     6.220$ & $     0.060$ & $+     0.140/-     0.050$ &$+     0.140/-     0.030$& $+     0.069/-     0.069$ & $+     0.218/-     0.108$\\ 
$\sigma_{W^-}$ $\times BR(W^- \to \ell^-\nu)$ [nb] & $     4.320$ & $     0.030$ & $+     0.100/-     0.030$ &$+     0.110/-     0.020$& $+     0.039/-     0.039$ & $+     0.157/-     0.061$\\ 
$\sigma_{Z}$ $\times BR(Z^0\to\ell^+\ell^-)$ [nb] & $     0.980$ & $     0.010$ & $+     0.025/-     0.007$ &$+     0.025/-     0.006$& $+     0.012/-     0.012$ & $+     0.039/-     0.018$\\ 
$\sigma_{t\bar{t}}$ [pb] & $     147.31$ & $      4.10$& $+      1.74/-      3.09$& $+      2.47/-     12.71$&$+      0.98/-      1.24$ &$+      5.18/-     13.76$\\ 
$\sigma_{gg\to H}(120\mbox{GeV})$ [pb] & $      11.79$ & $      0.24$& $+      0.04/-      0.18$& $+      0.13/-      0.75$&$+      0.32/-      0.32$ &$+      0.42/-      0.88$\\ 
$\sigma_{gg\to H}(180\mbox{GeV})$ [pb] & $       4.86$ & $      0.12$& $+      0.02/-      0.08$& $+      0.07/-      0.37$&$+      0.12/-      0.12$ &$+      0.17/-      0.41$\\ 
$\sigma_{gg\to H}(240\mbox{GeV})$ [pb] & $       2.57$ & $      0.07$& $+      0.01/-      0.05$& $+      0.04/-      0.22$&$+      0.05/-      0.05$ &$+      0.09/-      0.25$\\ 
\hline
\end{tabular}
\end{center}
\end{scriptsize}
Table 9: Benchmark cross section predictions and uncertainties for HERAPDF1.0 for $W^{\pm},Z, t\bar{t}$ and Higgs production (120, 180, 240 GeV) at 7 TeV. The central prediction is given in column 2. Errors are quoted at the 68\% c.l. The column labeled Exp. errors stands for the symmetric
experimental uncertainty, Model error --- for the asymmetric model uncertainty,
Param. error --- for the asymmetric parameterization uncertainty and $\alpha_S$
error for the uncertainty due to the $\Delta \alpha_S=0.002$ variation.
The Total error stands for the total uncertainty calculated by adding the negative
and positive variations in quadrature. Higgs boson cross sections have been corrected for the finite top mass effect (a factor of 1.06 for 120 GeV, 1.15 for 180 GeV and 1.31 for 240 GeV).

\begin{center}
\begin{tabular}{|c||c|c|c|c|c|c|}
\hline
 $y_W$ & $\frac{d \sigma_{W^+}}{dy} \times BR$ & PDF  $+\alpha_S$ Error &
       $\frac{d \sigma_{W^-}}{dy} \times BR$ & PDF  $+\alpha_S$ Error &
       $\frac{d \sigma_{Z^0}}{dy} \times BR$ & PDF  $+\alpha_S$ Error \\
\hline
$      0.0 $ & $     0.878$ & $^{+     0.039}_{-     0.021}$ & $     0.771$ & $^{+     0.025}_{-     0.012}$   & $    0.1634$ & $^{+    0.0074}_{-    0.0035}$ \\
$      0.4 $ & $     0.880$ & $^{+     0.040}_{-     0.020}$ & $     0.765$ & $^{+     0.024}_{-     0.012}$   & $    0.1625$ & $^{+    0.0077}_{-    0.0035}$ \\
$      0.8 $ & $     0.885$ & $^{+     0.041}_{-     0.019}$ & $     0.748$ & $^{+     0.025}_{-     0.011}$   & $    0.1606$ & $^{+    0.0076}_{-    0.0033}$ \\
$      1.2 $ & $     0.887$ & $^{+     0.039}_{-     0.016}$ & $     0.720$ & $^{+     0.026}_{-     0.010}$   & $    0.1568$ & $^{+    0.0077}_{-    0.0029}$ \\
$      1.6 $ & $     0.892$ & $^{+     0.039}_{-     0.014}$ & $     0.680$ & $^{+     0.027}_{-     0.010}$   & $    0.1514$ & $^{+    0.0070}_{-    0.0025}$ \\
$      2.0 $ & $     0.893$ & $^{+     0.032}_{-     0.015}$ & $     0.625$ & $^{+     0.024}_{-     0.011}$   & $    0.1438$ & $^{+    0.0057}_{-    0.0030}$ \\
$      2.4 $ & $     0.875$ & $^{+     0.033}_{-     0.020}$ & $     0.548$ & $^{+     0.023}_{-     0.013}$   & $    0.1313$ & $^{+    0.0056}_{-    0.0032}$ \\
$      2.8 $ & $     0.818$ & $^{+     0.047}_{-     0.025}$ & $     0.436$ & $^{+     0.021}_{-     0.015}$   & $    0.1094$ & $^{+    0.0081}_{-    0.0037}$ \\
$      3.2 $ & $     0.658$ & $^{+     0.055}_{-     0.029}$ & $     0.291$ & $^{+     0.025}_{-     0.016}$   & $    0.0765$ & $^{+    0.0090}_{-    0.0033}$ \\
$      3.6 $ & $     0.380$ & $^{+     0.043}_{-     0.024}$ & $     0.135$ & $^{+     0.030}_{-     0.011}$   & $    0.0337$ & $^{+    0.0066}_{-    0.0022}$ \\
$      4.0 $ & $     0.090$ & $^{+     0.014}_{-     0.009}$ & $     0.028$ & $^{+     0.019}_{-     0.005}$   & $    0.0048$ & $^{+    0.0033}_{-    0.0009}$ \\

$      4.4 $ & $     0.002$ & $^{+     0.004}_{-     0.001}$ & $     0.001$ & $^{+     0.004}_{-     0.001}$   & $    0.0000$ & $^{+    0.0002}_{-    0.0000}$ \\
\hline
\end{tabular}
\end{center}
Table 10: Benchmark cross section predictions ($d\sigma/dy \times BR$ in nb) for HERAPDF1.0 set calculated at 
$\alpha_S=0.1176$ as a function of boson rapidity. All sources of error calculated above are included. 

\subsubsection{MSTW2008}

In the following sub-section, the tables of relevant cross sections for the MSTW2008 PDFs are given (Tables 11-14). The predictions for the central value of $\alpha_s(m_Z)$ are given in bold. 


  \begin{center}    
    \begin{tabular}{|l||c|c|c|}
      \hline
      {\small $\alpha_s(m_Z)$ }& {\small $\sigma_{W^+}*BR(W^+\rightarrow l^+\nu$)[nb]} &{\small $\sigma_{W^-}*BR(W^-\rightarrow l^-\nu)[nb]$ } & {\small $\sigma_{Z^o}*BR(Z^o\rightarrow l^+ l^-)[nb]$}   \\
	\hline
0.1187 &  5.897 & 4.150 & 0.9336  \\
0.1194 &  5.927 & 4.171 & 0.9398  \\
$\bf 0.1202$ &  $\bf 5.957$ & $\bf 4.190 $ & $\bf 0.9442 $  \\
0.1208 &  5.982 & 4.208 & 0.9479  \\
0.1214 &  6.008 & 4.225 & 0.9516  \\
0.1190 &  5.911 & 4.160 & 0.9374  \\
      \hline
    \end{tabular}
  \end{center}
Table 11: Benchmark cross section predictions for MSTW 2008 for $W^{\pm},Z$ and $t\bar{t}$ production at 7 TeV, as a function of $\alpha_s(m_Z)$. The results for the central value of $\alpha_s(m_Z)$ for MSTW 2008 (0.1202) are shown in bold.  


  \begin{center}    
    \begin{tabular}{|l||c|c|c|c|}
      \hline
      $\alpha_s(m_Z)$ & $\sigma_{gg\rightarrow Higgs}(120 ~GeV)[pb]$ & $\sigma_{gg\rightarrow Higgs}(180 ~GeV)[pb]$& $\sigma_{gg\rightarrow Higgs}(240 ~GeV)[pb]$ & {\small $\sigma_{t\bar{t}}[pb]$}\\
	\hline
0.1187 & 12.13  & 5.08  & 2.74  & 163.5\\
0.1194 & 12.27  & 5.14  & 2.77 & 165.8\\
$\bf 0.1202$ &  $\bf 12.41$ & $\bf 5.19 $ & $\bf 2.81 $ & $ \bf 168.1 $  \\
0.1208 & 12.53  & 5.24 & 2.83 & 170.0 \\
0.1214 & 12.64  & 5.29 & 2.86 & 171.9  \\
0.1190 & 12.18  & 5.10 & 2.76 & 164.4  \\
      \hline
    \end{tabular}
  \end{center}
Table 12: Benchmark cross section predictions for MSTW 2008 for $gg\rightarrow Higgs$ production (masses of 120, 180 and 240 GeV), and for $t\bar{t}$ production,  at 7 TeV, as a function of $\alpha_s(m_Z)$. The results for the central value of $\alpha_s(m_Z)$ for MSTW 2008 (0.1202) are shown in bold.  Cross sections have been corrected for the finite top mass effect (a factor of 1.06 for 120 GeV, 1.15 for 180 GeV and 1.31 for 240 GeV). 

  \begin{center}    
    \begin{tabular}{|l||c|c|c|c|c|c|}
      \hline
{\small Process} & {\small $\sigma$} & {\small PDF (asym)} & {\small PDF (sym)} & {\small $\alpha_s(m_Z)$ error} 
& combined \\
	\hline
{\footnotesize $\sigma_{W^+}*BR(W^+\rightarrow l^+\nu)[nb]$}  &  5.957 & +0.129/-0.097 & 0.107 & +0.051/-0.060 & +0.145/-0.121 \\
{\footnotesize $\sigma_{W^-}*BR(W^-\rightarrow l^-\nu)[nb]$ } &  4.190 & +0.092/-0.071 & 0.079 & +0.035/-0.040 & +0.104/-0.080 \\
{\footnotesize $\sigma_{Z^o}*BR(Z^o\rightarrow l^+ l^-)[nb]$}  &  0.944  &  +0.020/-0.014 &  0.017 & +0.007/-0.009 & +0.023/-0.0018 \\
$\sigma_{t\bar{t}}[pb]$ &  168.1 & +4.7/-5.6 & 4.9  & +3.8/-4.6 & +7.2/-6.0 \\
{\footnotesize $\sigma_{gg\rightarrow Higgs}(120~ GeV)[pb]$ } &  12.41 & +0.17/-0.21 & 0.19 & +0.23/-0.28 & +0.40/-0.34 \\
{\footnotesize $\sigma_{gg\rightarrow Higgs}(180~ GeV)[pb]$ } &  5.194 & +0.090/-0.106 & 0.095 & +0.094/-0.115 & +0.177/-0.136\\
{\footnotesize $\sigma_{gg\rightarrow Higgs}(240~ GeV)[pb]$}  &  2.806 & +0.057/-0.069 & 0.062 & +0.052/-0.063 & +0.101/-0.077 \\
\hline
\end{tabular}
\end{center}
Table 13: Benchmark cross section predictions and uncertainties for MSTW 2008 for $W^{\pm},Z, t\bar{t}$ and Higgs production (120, 180, 240 GeV) at 7 TeV. The central prediction is given in column 2. Errors are quoted at the 68\% c.l. The symmetric and asymmetric forms for the PDF errors are given. The $\alpha_s$ uncertainty is the deviation of the central prediction at the upper and lower 68\% c.l. values
of $\alpha_s(M_Z^2)$. The combination follows the procedure outlined in the text.  

  \begin{center}    
    \begin{tabular}{|l||c|c|c|c|c|c|c|}
      \hline
      $ y_W $ & ${d\sigma_{W^+}\over{dy}}*BR$ & PDF + $\alpha_s$ Error & ${d\sigma_{W^-}\over{dy}}*BR$ & PDF + $\alpha_s$ Error &${d\sigma_{Z^o}\over{dy}}*BR$ & PDF + $\alpha_s$ Error \\
\hline
-4.4 &  0.0024 & 0.0001 & 0.00007 & 0.00003 & 0.000011 & 0.000001 \\
-4.0 &  0.087 & 0.003 & 0.014 & 0.002 & 0.0037 & 0.0001 \\
-3.6 &  0.353 & 0.017 & 0.113 & 0.006 & 0.029 & 0.001  \\
-3.2 &  0.616 & 0.021 & 0.276 & 0.011 & 0.070 & 0.002 \\
-2.8 &  0.787 & 0.022 & 0.422 & 0.015 & 0.105 & 0.003 \\
-2.4 &  0.863 & 0.023 & 0.523 & 0.015 & 0.128 & 0.004 \\
-2.0 &  0.888 & 0.022 & 0.608 & 0.017 & 0.141 & 0.003 \\
-1.6 &  0.879 & 0.020 & 0.678 & 0.017 & 0.151 &  0.004 \\
-1.2 &  0.860 & 0.019 & 0.716 & 0.020 & 0.154 & 0.004 \\
-0.8 &  0.843 & 0.027 & 0.736 & 0.018 & 0.157 & 0.005 \\
-0.4 &  0.856 & 0.023 & 0.770 & 0.019 & 0.161 & 0.005 \\
0.0 &  0.829  & 0.022 & 0.760 & 0.020 & 0.159 & 0.004  \\
0.4 &  0.834 & 0.023 & 0.765 & 0.021 & 0.161 & 0.005  \\
0.8 &  0.855 & 0.022 & 0.742 & 0.019 & 0.157 & 0.004  \\
1.2 &  0.865 & 0.026 & 0.719 & 0.018 & 0.154 & 0.004 \\
1.6 &  0.875 & 0.021 & 0.677 & 0.021 & 0.151 & 0.003  \\
2.0 &  0.890 & 0.026 & 0.608 & 0.018 & 0.142 & 0.004  \\
2.4 &  0.866 & 0.020 & 0.530 & 0.015 & 0.129 & 0.003 \\
2.8 &  0.798 & 0.020 & 0.413 & 0.014 & 0.104 & 0.003 \\
3.2 &  0.611 & 0.019 & 0.280 & 0.012 & 0.070 & 0.002  \\
3.6 &  0.341 & 0.017 & 0.112 & 0.008 & 0.029 & 0.001  \\
4.0 &  0.096 & 0.003 & 0.015 & 0.002 & 0.0042 & 0.0002  \\
4.4 &  0.0022 & 0.0001 & 0.00008 & 0.00003  & 0.000010 & 0.000001  \\
\hline

\end{tabular}
\end{center}
Table 14: Benchmark cross section predictions ($d\sigma/dy*BR$ in nb) for MSTW 2008 for $W^{\pm},Z^o$ production at 7 TeV, as a function of boson rapidity. 

\subsubsection{NNPDF2.0}

We show results for the NNPDF2.0 PDF set, with the default
NNPDF choice of $\alpha_s(m_Z)=0.119$, as well as for other
values of $\alpha_s(m_Z)$ in Table 15
for all
the benchmark LHC observables. Note that for each value
of $\alpha_s$ we provide the central prediction and the 
associated PDF uncertainties. For the combined PDF+$\alpha_s$
uncertainty, we assume the benchmark value of 
$\alpha_s\lp m_Z\rp=0.1190\pm 0.0012$ as a 1--sigma uncertainty.

Next in Table 16
we provide
for  the same observables the 
benchmark cross section predictions and uncertainties for the NNPDF2.0
set with the combined PDF and strong coupling uncertainties 
at LHC 7 TeV. 
In the last column,
the PDF and $\alpha_s$  errors are combined using exact error propagation
as discussed in Sect.~\ref{sec:NNPDFdetails}.

Finally in Table 17
we provide
the benchmark cross section predictions for the
differential rapidity distributions
 ($d\sigma/dy\cdot$BR in pb) for NNPDF2.0 for $W^{\pm}$ and $Z^0$ 
production at 7 TeV, as a function of the
boson rapidity. We provide for each bin in rapidity the combined PDF
and $\alpha_s$ uncertainty using exact error propagation.

  \begin{center}
  {\small
  \begin{tabular}{|c|c|c|}
    \hline
        & $\sigma(W^+){\rm Br}\lp W^+ \to l^+\nu_l\rp\,$ 
         & $\sigma(W^-){\rm Br}\lp W^- \to l^+\nu_l\rp\,$
 \\
          \hline
\hline
    $\alpha_s$=0.115  & $ 5.65\pm 0.13 $ nb 
& $3.86\pm 0.09$ nb \\
$\alpha_s$=0.117  & $ 5.73 \pm 0.13$ nb 
& $3.91\pm 0.08 $ nb  \\
\bf $\alpha_s$=0.119  & \bf $ 5.80\pm 0.15 $ nb 
\bf & $3.97\pm 0.09 $ nb \\
$\alpha_s$=0.121  & $ 5.87 \pm 0.13 $ nb 
& $ 4.03\pm 0.08$ nb \\
$\alpha_s$=0.123  & $ 5.98\pm 0.14 $ nb 
& $4.10\pm 0.10$ nb \\
\hline
  \end{tabular}

$\quad$\\\vspace{1cm}  \begin{tabular}{|c|c|c|}
    \hline
        & $\sigma(Z^0){\rm Br}\lp Z^+ \to l^+l^-\rp\,$ 
     
& $\sigma(t\bar{t})$ \\
          \hline
\hline
    $\alpha_s$=0.115  & $ 886\pm 18 $ pb 
  & $ 156 \pm 5$ pb  \\
$\alpha_s$=0.117  & $ 898\pm 16 $ pb 
 & $ 162 \pm 5 $  pb   \\
\bf $\alpha_s$=0.119  &\bf $ 909\pm 19 $ pb 
 &\bf $ 169 \pm 6 $ pb \\
$\alpha_s$=0.121  & $  921\pm 17  $ pb 
& $ 176 \pm 6 $ pb \\
$\alpha_s$=0.123  & $ 937\pm 21 $ pb 
 & $182 \pm 7 $  pb \\
\hline
  \end{tabular}
$\quad$\\\vspace{1cm}
  \begin{tabular}{|c|c|c|c|}
    \hline
        & $\sigma(H)(120\,{\rm GeV})$ 
         & $\sigma(H)(180\,{\rm GeV})$ 

     & $\sigma(H)(240\,{\rm GeV})$ \\
          \hline
\hline
    $\alpha_s$=0.115  & $  11.61\pm 0.25 $ pb 
& $ 4.86\pm 0.12 $ pb & $2.69\pm 0.066  $ pb  \\
$\alpha_s$=0.117  & $ 11.90\pm 0.19 $ pb 
& $ 5.05\pm 0.09 $ pb &
 $ 2.75\pm 0.066  $ pb  \\
\bf $\alpha_s$=0.119  &\bf $ 12.30\pm 0.18  $ pb
 & \bf $ 5.22\pm 0.10  $ pb 
& $2.84\pm 0.066$ pb \\
$\alpha_s$=0.121  & $ 12.66\pm 0.18 $ pb 
& $ 5.38\pm 0.09  $ pb 
& $ 2.93\pm 0.066 $ pb \\
$\alpha_s$=0.123  & $ 12.92\pm 0.20 $ pb 
& $ 5.49\pm 0.10  $ pb 
& $ 3.00\pm 0.079  $ pb \\
\hline
  \end{tabular}
}
\end{center}
Table 15: Benchmark cross section predictions for NNPDF2.0
 for $W^{\pm}$, $Z^0$, $t\bar{t}$ and Higgs production at 7 TeV, as a function of $\alpha_s(m_Z)$. The results
for the central value of $\alpha_s(m_Z)$ for NNPDF2.0 (0.119) 
are shown in bold. For each value of $\alpha_s(m_Z)$ we provide both
the central prediction and the associated 1--sigma PDF uncertainties. The Higgs boson cross sections have been corrected for the finite top mass effect (1.06,1.15,1.31) for the three values of the Higgs boson mass. 

  \begin{center}
  \footnotesize
  \begin{tabular}{|c|c|c|c|c|}
    \hline
     Process   & Cross section & PDF errors (1--$\sigma$)
& $\alpha(m_Z)$ error &
  PDF+$\alpha_s$ error\\
          \hline
\hline
$\sigma(W^+){\rm Br}\lp W^+ \to l^+\nu_l\rp\,$  [nb] & 5.80 & 0.15 & 0.04 & 0.16  \\
$\sigma(W^+){\rm Br}\lp W^- \to l^+\nu_l\rp\,$  [nb] & 3.97 & 0.09 & 0.04& 0.10  \\
 $\sigma(Z^0){\rm Br}\lp Z^+ \to l^+l^-\rp\,$ [nb]  & 0.909 & 0.022 & 0.007& 0.023 \\
$\sigma(t\bar{t})$ [pb]  & 169 & 6 & 4 & 7 \\
$\sigma(H)(120\,{\rm GeV})$ [pb]  & 12.30 & 0.18 & 0.23 & 0.29 \\
$\sigma(H)(180\,{\rm GeV})$ [pb]  & 5.22 & 0.10 & 0.10& 0.14 \\
$\sigma(H)(240\,{\rm GeV})$ [pb]  & 2.84 & 0.066 & 0.052& 0.092 \\
\hline
\end{tabular}
\end{center}
Table 16: Benchmark cross section predictions and uncertainties for NNPDF2.' for $W^{\pm}$, $Z^0$, $t\bar{t}$ and Higgs production (120, 180, 240 GeV)
at LHC 7 TeV. The central prediction is given in column 2. We provide
PDF uncertainties as 1--sigma
uncertainties.
In the last column,
the PDF and $\alpha_s$  errors are combined using exact error propagation
as discussed in Sect.~\ref{sec:NNPDFdetails}. The Higgs boson cross sections have been corrected for the finite top mass effect (1.06,1.15,1.31) for the three values of the Higgs boson mass. 

  \begin{center}
  \scriptsize
  \begin{tabular}{|c|c|c|c|c|c|c|}
\hline
$y_{W/Z}$ 
 & $d\sigma_{W^+}/dy\cdot$BR & PDF+$\alpha_s$ error &
  $d\sigma_{W^-}/dy\cdot$BR & PDF+$\alpha_s$ error &
  $d\sigma_{Z^0}/dy\cdot$BR & PDF+$\alpha_s$ error \\

\hline
\hline
   -4.40 &   0.0024 &   0.0003 &   0.0023 &   0.00025   &    0.00001 & 0.00001\\
   -4.00 &   0.090 &    0.0048 &   0.021 &    0.0021  &     0.0039 &  0.00032\\
   -3.60 &   0.352 &    0.020 &    0.122 &    0.007  &      0.030 &   0.002\\
   -3.20 &   0.588 &    0.023 &    0.279 &    0.014  &      0.070 &   0.0037\\
   -2.80 &   0.771 &    0.030 &    0.395 &    0.016   &     0.102 &   0.0042\\
   -2.40 &   0.842 &    0.030 &    0.500 &    0.020   &     0.124 &   0.0043\\
   -2.00 &   0.862 &    0.029 &    0.569 &    0.018   &     0.136 &   0.0053\\
   -1.60 &   0.851 &    0.025 &    0.624 &    0.020  &      0.144 &   0.0038\\
   -1.20 &   0.827 &    0.040 &    0.655 &    0.015  &      0.147 &   0.0039\\
   -0.80 &   0.831 &    0.020 &    0.687 &    0.021  &      0.150 &   0.0045\\
   -0.40 &   0.819 &    0.023 &    0.692 &    0.017  &      0.151 &   0.0034\\
    0.00 &   0.815 &    0.020 &    0.701 &    0.019 &       0.153 &   0.0049\\
    0.40 &   0.836 &    0.024 &    0.713 &    0.016 &       0.154 &   0.0031\\
    0.80 &   0.820 &    0.023 &    0.678 &    0.016 &       0.149 &   0.0046\\
    1.20 &   0.840 &    0.026 &    0.667 &    0.017 &       0.149 &   0.0044\\
    1.60 &   0.858 &    0.029 &    0.623 &    0.020 &       0.145 &   0.0039\\
    2.00 &   0.860 &    0.029 &    0.583 &    0.024 &       0.140 &   0.0051\\
    2.40 &   0.861 &    0.029 &    0.508 &    0.019 &       0.126 &   0.0046\\
    2.80 &   0.771 &    0.032 &    0.397 &    0.017 &       0.099 &   0.0047\\
    3.20 &   0.598 &    0.024 &    0.260 &    0.013 &       0.066 &   0.0037\\
    3.60 &   0.338 &    0.017 &    0.119 &    0.007 &       0.030 &   0.0025\\
    4.00 &   0.076 &    0.004 &    0.018 &    0.0023 &      0.0044 &  0.00037\\
    4.40 &   0.0016 &   0.0007 &   0.0031 &   0.00031 &  0.00001 &    0.00\\
\hline
\end{tabular}
\end{center}
Table 17: Benchmark cross section predictions ($d\sigma/dy\cdot$BR in nb) for NNPDF2.0 for $W^{\pm}$ and $Z^0$ production at 7 TeV, as a function of
boson rapidity. We provide for each bin in rapidity the combined PDF
and $\alpha_s$ uncertainty using exact error propagation.


\subsection{Comparison of $W^+,W^-,Z^o$ rapidity distributions}

\begin{figure}[ht]
\begin{center}
\includegraphics[height=95mm,angle=0]{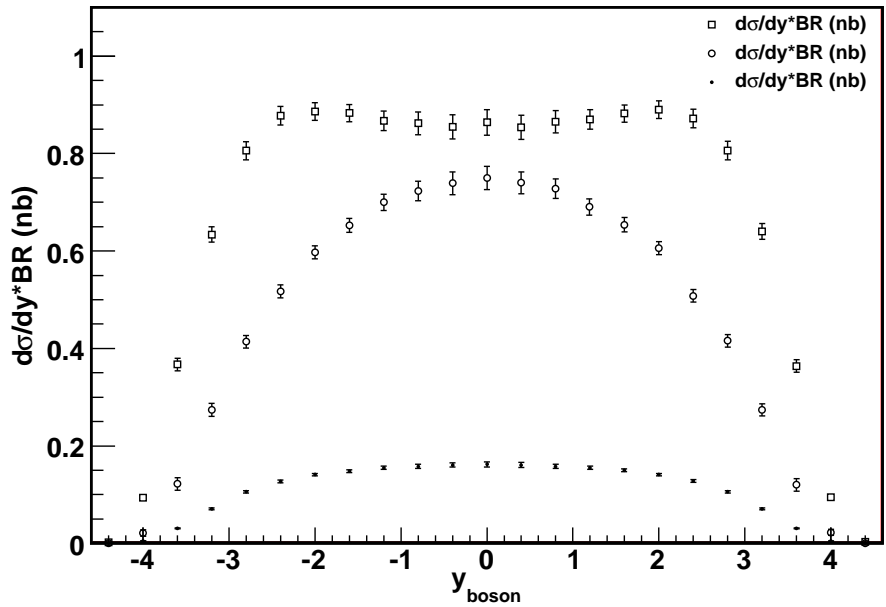}
\caption{\label{fig:W_cteq66} The $W^+, W^-$ and $Z$ cross sections at 7 TeV (along with the PDF uncertainties) as a function of rapidity for the CTEQ6.6 PDFs. The error bars indicate the PDF+$\alpha_s$ uncertainties shown in Table 6.}
\end{center}
\end{figure}

NLO predictions for the $W^+, W^-$ and $Z$ cross sections (along with the PDF uncertainties) are plotted as a function of the boson's rapidity are plotted in Figure~\ref{fig:W_cteq66} using the CTEQ6.6 PDFs. 

In the figures below, a comparison is made for the predictions of the $W^+,W^-,Z^0$ boson rapidity distributions for the PDFs discussed in this note. In general, the predictions are in reasonable agreement, but differences can be observed that should be detectable with a reasonably-sized data sample at 7 TeV.

\begin{figure}[h]
\begin{center}
\includegraphics[height=95mm,angle=0]{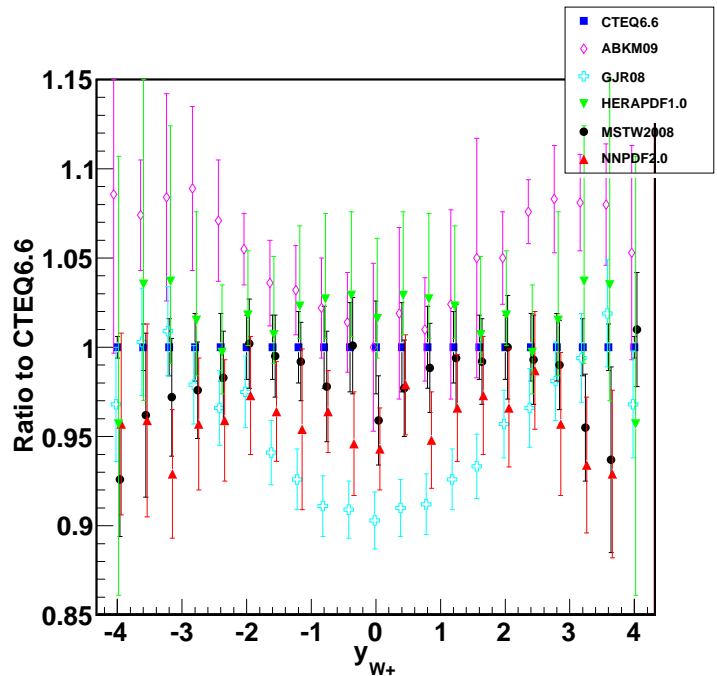}
\caption{\label{fig:wprap} The $W^+$ cross section at 7 TeV as a function of rapidity for the PDFs discussed in this note, normalized to the result of CTEQ6.6. The error bars indicate the PDF(+$\alpha_s$/model/...) uncertainties as defined in the tables provided by each PDF group.}
\end{center}
\end{figure}

\begin{figure}[h]
\begin{center}
\includegraphics[height=90mm,angle=0]{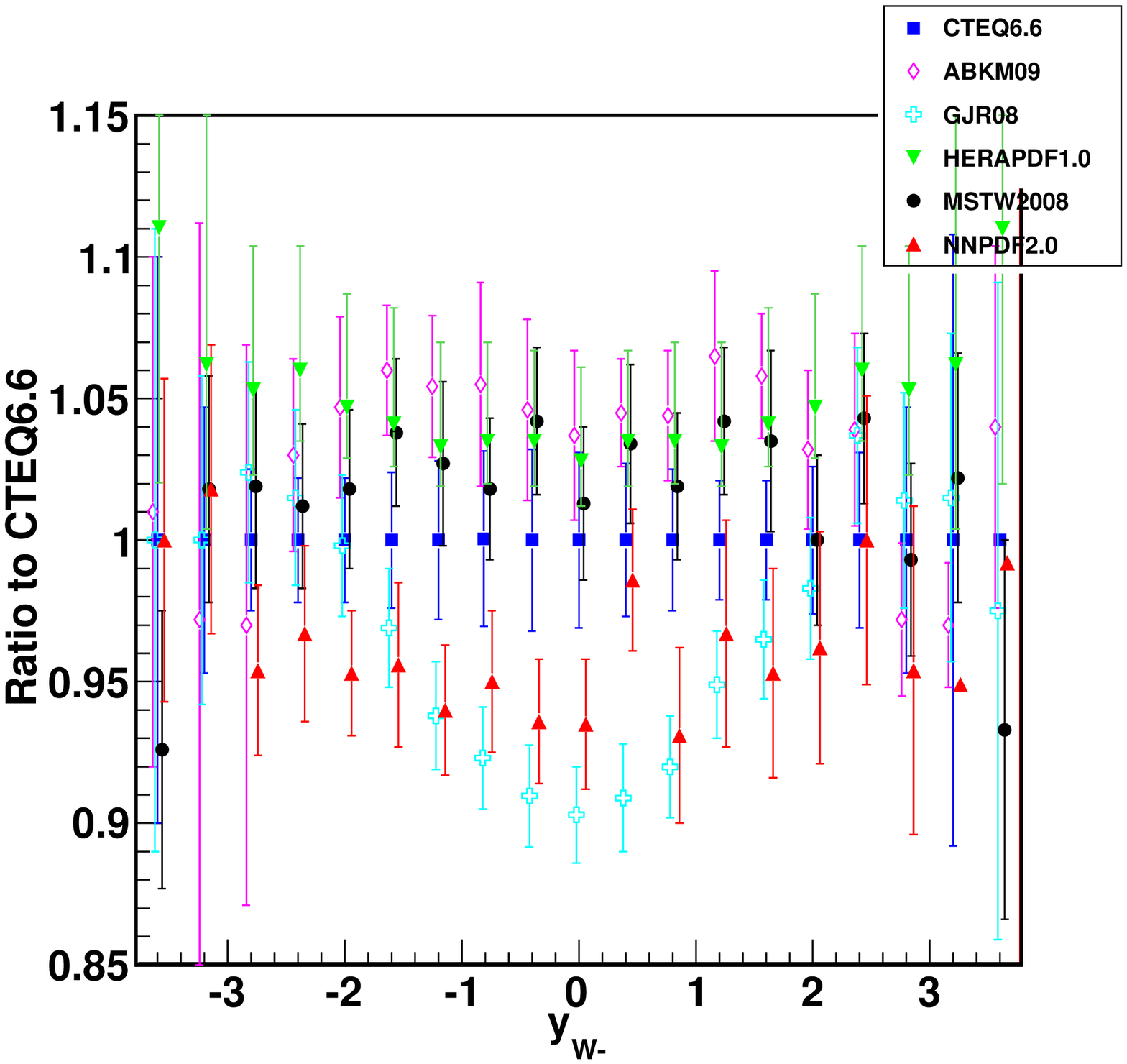}
\caption{\label{fig:wmrap} The $W^-$ cross section at 7 TeV as a function of rapidity for the PDFs discussed in this note, normalized to the result of MSTW2008. The error bars indicate the PDF(+$\alpha_s$/model/...) uncertainties as defined in the tables provided by each PDF group.}
\end{center}
\end{figure}

\begin{figure}[h]
\begin{center}
\includegraphics[height=90mm,angle=0]{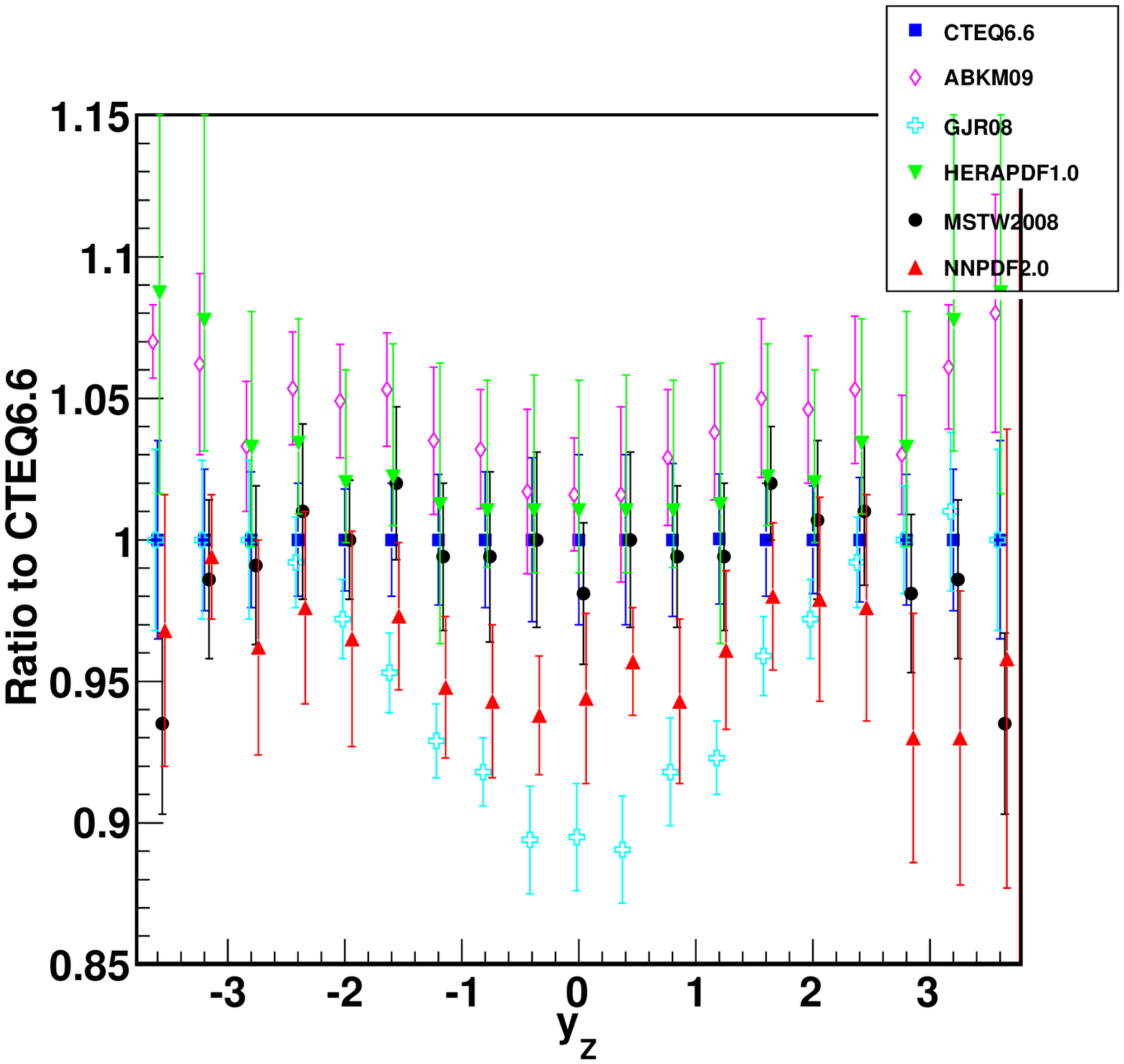}
\caption{\label{fig:zrap} The $Z^o$ cross section at 7 TeV as a function of rapidity for the PDFs discussed in this note, normalized to the result of MSTW2008. The error bars indicate the PDF(+$\alpha_s$/model/...) uncertainties as defined in the tables provided by each PDF group.}
\end{center}
\end{figure}


\section{Summary}
\label{sec:summary}

In this interim report, we have tried to provide a snapshot of our current understanding of PDFs and the associated experimental and theoretical uncertainties, and of predictions for benchmark cross sections at the LHC (7 TeV) and their corresponding uncertainties. This snapshot will be updated as new input data/theoretical treatments become available. Many of the PDFs discussed in this note are now not the most recent generation from the respective PDF groups, but we have concentrated on these since they are in the most common use by the experimental groups.

\clearpage


\end{document}